\documentclass[fleqn,usenatbib]{mnras}

\usepackage{newtxtext,newtxmath, amsmath}

\usepackage[T1]{fontenc}
\usepackage{ae,aecompl}
\usepackage[normalem]{ulem}

\usepackage{graphicx}	
\usepackage{amsmath}	
\usepackage{amssymb}	
\usepackage[table]{xcolor}

\usepackage{array}
\newcolumntype{L}[1]{>{\raggedright\let\newline\\\arraybackslash\hspace{0pt}}m{#1}}
\newcolumntype{C}[1]{>{\centering\let\newline\\\arraybackslash\hspace{0pt}}m{#1}}
\newcolumntype{R}[1]{>{\raggedleft\let\newline\\\arraybackslash\hspace{0pt}}m{#1}}

\usepackage[normalem]{ulem}

\newcommand{\Reff}{R_{\rm e}}	
\newcommand{\Mgc}{M_{\rm GC}}	
\newcommand{\MHI}{M_{\rm HI}}	
	
\newcommand{\Mtwo}{M_{\rm 200}}	
\newcommand{\Mdyn}{M_{\rm dyn}}	
\newcommand{\Mdyndm}{M_{\rm dyn, DM}}	
\newcommand{\Msun}{{\rm M}_\odot}	
\newcommand{\Lsun}{{\rm L}_\odot}	
	
\newcommand{\Ngc}{N_{\rm GC}}

\title[GC system--halo mass relation]{Extending the Globular Cluster System-Halo Mass Relation to the Lowest Galaxy Masses}

\author[D. A. Forbes et al.]{
Duncan A. Forbes$^{1}$\thanks{E-mail: dforbes@swin.edu.au}, Justin I. Read$^2$, Mark Gieles$^2$, Michelle L. M. Collins$^2$
\\
$^{1}$Centre for Astrophysics \& Supercomputing, Swinburne University, Hawthorn VIC 3122, Australia\\
$^{2}$Department of Physics, University of Surrey, Guildford GU2 7XH, UK\\
}

\date{Accepted XXX. Received YYY; in original form ZZZ}

\pubyear{2015}

\begin{document}
\label{firstpage}
\pagerange{\pageref{firstpage}--\pageref{lastpage}}
\maketitle

\begin{abstract}
High mass galaxies, with halo masses $M_{200}\ge 10^{10}\,\Msun$, reveal a remarkable near-linear relation between their globular cluster (GC) system mass and their host galaxy halo mass. Extending this relation to the mass range of dwarf galaxies has been problematic due to the difficulty in measuring independent halo masses. Here we derive new halo masses based on stellar and HI gas kinematics for a sample of nearby dwarf galaxies with GC systems. We find that the GC system mass--halo mass relation for galaxies populated by GCs holds from halo masses of $M_{200}$ $\sim$ 10$^{14}$\,$\Msun$ down to below $M_{200}$  $\sim 10^9$\,$\Msun$, although there is a substantial increase in scatter towards low masses. 
In particular, three well-studied ultra diffuse galaxies, with dwarf-like stellar masses, reveal a wide range in their GC-to-halo mass ratios. We compare our GC system--halo mass relation to the recent model of El Badry et al., finding that their fiducial model does not reproduce our data in the low mass regime. This may suggest that GC formation needs to be more efficient than assumed in their model, or it may be due to the onset of stochastic GC occupation in low mass halos. Finally, we briefly discuss the stellar mass-halo mass relation for our low mass galaxies with GCs, and we suggest some nearby dwarf galaxies for which searches for GCs may be fruitful.
\end{abstract}

\begin{keywords}
galaxies: dwarf -- galaxies: halos  -- galaxies: star clusters
\end{keywords}










\begin{table*}
\caption{Pressure-supported Local Group dwarfs with globular clusters and velocity dispersion measurements. 
}
\label{tab:way}
\begin{tabular}{llccccccccc}
\hline
Name & Type & $M_V$ & $M_*$ & $\MHI$ & $\Reff$ & $\sigma$ & log $M_{200}$ & $\Ngc$ & $\Mgc$ & GC Ref \\
     & & [mag] & [10$^7$ $\Msun$] & [10$^7$ $\Msun$] & [pc] & [km/s] & [$\Msun$] &   & [10$^4$ $\Msun$] & \\
     \hline
Sagittarius & dSph & -15.5 & 13.2 & -- & 2587 (219) & 11.4 (0.7) & 10.78 & 8 & 278 & L10\\
Fornax & dSph & -13.4 & 2.0 & 0.017 & 710 (70) & 11.8 (0.2) & 9.06 & 5 & 73 & M03\\
Eridanus II & dSph & -7.1 & 0.0059 &  -- & 280 (28) & 6.9 (1.1) & 9.05 & 1 & 0.4 & Cr16\\
\hline
NGC~205 & dE & -16.5 & 34.0 & 0.040 & 590 (31) & 35.0 (5.0) & 9.83 & 11 & 140 & D88\\
NGC~185 & dE & -14.8 & 7.1 & 0.011 & 458 (45) & 24.0 (1.0) & 10.64 & 8 & 68 & V13\\
NGC~147 & dE & -14.6 & 5.9 & -- & 623 (62) & 16.0 (1.0) & 9.27  & 10 & 130 & V13\\
And I & dSph & -11.7 & 0.40 & -- & 672 (69) & 10.6 (1.1) & 9.28 & 1 & 0.37 & C17\\
And XXV & dSph & -9.7 & 0.065 & -- & 642 (75) & 3.0 (1.0) & 7.12 & 1 & 1.5  & Cu16\\
\hline
\end{tabular}
\\\vspace{0.1cm}
Notes: The top half of the table lists dwarfs associated with the Milky Way and the lower half with the Andromeda galaxy. Absolute magnitude ($M_V$), stellar mass ($M_{\ast}$), HI gas mass ($\MHI$), half light radius ($\Reff$) and central velocity dispersion ($\sigma$) are taken from the compilation of \citet{2012AJ....144....4M} and online updates. $M/L_V = 1$ is used to calculate stellar mass. For Sagittarius see text for details. For Eridanus~II data comes from  from the discovery paper. 
The number of GCs ($\Ngc$) is taken from the GC reference; 
L10 = \citet{2010ApJ...718.1128L}, M03 = \citet{2003MNRAS.340..175M}, Cr16 = \citet{2016ApJ...824L..14C}, D88 = \citet{1988ApJ...334..159D}, V13 = \citet{2013MNRAS.435.3654V}, C17 = \citet{2017PASA...34...39C}, Cu16 = \citet{2016ApJ...829...26C}. 
See text for details of how the GC system mass ($\Mgc$) and halo mass ($\Mtwo$) are calculated. The online version of this table includes errors on the halo mass. 
\end{table*}

\section{Introduction}

The total mass of a globular cluster (GC) system reveals a remarkable, near linear, correlation with the total mass of its host galaxy halo over 4 decades in galaxy mass and which is largely insensitive to galaxy type or environment \citep{2009MNRAS.392L...1S,2010MNRAS.406.1967G,2014ApJ...787L...5H,2015ApJ...806...36H}. This correlation 
indicates a close scaling between GC systems and galaxy dark matter halos and suggests that it was established during the initial phases of galaxy formation \citep[e.g.][]{2009ApJ...706L.192B}, with subsequent evolution changing both GC and halo mass in roughly equal proportions \citep[e.g.][]{2005ApJ...623..650K}.

Based on weak lensing results, \citet{2009MNRAS.392L...1S} matched GC system masses ($\Mgc$) statistically to galaxy halo masses ($\Mtwo$) using their known host galaxy stellar masses ($M_{\ast}$). The resulting trend between GC system and halo mass was consistent with a linear relation for halo masses 
$10^{10}<\Mtwo/\Msun< 10^{14}$, with $\Mgc/M_{200} \simeq 5\times10^{-5}$. At higher masses an uncertain correction for intracluster GCs was needed. At lower masses, the $M_{\ast}-\Mtwo$ relation became very unreliable due to the difficulty of measuring halo masses with weak lensing. 
\citet{2009MNRAS.392L...1S} concluded that {\it ``...the correlation [of GC system mass and halo mass] for low galaxy masses must be confirmed with better halo mass estimates.''}

In subsequent work, \citet{2010MNRAS.406.1967G} assumed $M_{\ast}$ $\propto$ $M_{200}^{1.2}$ (based on the models of \citet{2006MNRAS.368....2D} for low-mass dwarf galaxies) extending the GC system relation to lower masses. Assuming this relations holds, they found early and late-type galaxies to have a similar relation, albeit with increased scatter at low masses. They suggested that the lowest mass galaxy that can, on average host a single GC, will have a luminosity of $M_V$ $\sim$ --9. 
This lower limit has since been surpassed with the discovery of the single old GC in the galaxy Eridanus II (which itself has a luminosity, $M_V = -7.1$, less than that of the typical Milky Way GC; \citealt{2016ApJ...824L..14C}). Its GC is roughly two orders of magnitude less massive than a typical Milky Way GC (i.e. $\sim2\times10^5\,\Msun$).

\citet{2014ApJ...787L...5H}, using weak lensing results from the CFHTLenS survey \citep{2015MNRAS.447..298H} and the database of GC systems from \citet{2013ApJ...772...82H}, were able to extend the relation down to stellar masses of $M_{\ast}\sim$10$^8$ M$_{\odot}$. They measured a GC system mass to halo mass ratio of 4 ($\times$ 10$^{-5}$) compared to 5.5 by  \citep{2010MNRAS.406.1967G} and 7 by \citep{2009MNRAS.392L...1S}. The recent study of \citet{2017ApJ...836...67H} extended the relation in halo mass but only added one galaxy (the Fornax dSph) with a halo mass below 10$^{10}$ $\Msun$. 

We also note the work of \citet{2016ApJ...826L...9Z} who focused on Eridanus II and its sole GC. On the assumption that the \citet{2010ApJ...717..379B} stellar mass--halo mass relation (i.e. $M_{\ast}$ $\propto$ $M_{200}^{1.4}$) extends down to the Eridanus II galaxy mass regime ($M_{\ast} = 6 \times 10^4 ~\Msun$), they concluded that its GC obeys the GC system mass-halo mass relation. 

Recently, \citet{2018arXiv180503652E} have followed the hierarchical assembly of GC systems using a semi-analytical model. Their model has no causal connection between GCs and dark matter. Instead, they suggest that a long history of galaxy mergers results in a linear GC system mass--halo mass relation for high mass systems 
At low halo masses they predict a non-linear relation with such systems {\it ``...retain[ing] the most information about the physical conditions under which GCs formed."}
Below $M_{200} \sim 3 \times10^{10}\,\Msun$ they predict most halos to have no GCs as the result of low  gas accretion rates and relatively high outflow rates.



Here we collect data on dwarf galaxies with known bona fide GCs (i.e. compact star clusters, older than several Gyr) in the Local Group. This is supplemented by old GCs in dwarf galaxies in nearby low-density environments. We use either the stellar velocity dispersion or the HI gas rotation to derive a total halo mass for individual galaxies. We also add up the masses of the individual GCs in each galaxy rather than simply using a mean mass multiplied by the number of GCs. Although this latter approach is reasonable for estimating the total GC system mass of massive 
galaxies \citep{2009MNRAS.392L...1S,2014ApJ...787L...5H} it would lead to significant errors for low-mass galaxies. 
We then revisit the GC system mass--halo mass relation. In Appendix A we list some nearby dwarf galaxies with GCs but no kinematic information; in Appendix B
nearby dwarfs with no reported GCs, but for which a detailed search for GCs may prove successful; and in Appendix C we show the GC number vs halo mass relation as an illustration of why GC masses need to be determined individually for low mass galaxies. 

\section{Local Group dwarf galaxies with GC systems}

The Local Group and its immediate surroundings contain over 100 galaxies \citep[e.g.][]{2012AJ....144....4M}. The vast majority are dwarf galaxies fainter than $M_V$ $\sim$ --18, with only a dozen or so reported to host a GC system.
In Table 1 we list the 7 Local Group dSph and dE galaxies (the most massive being NGC 205) that lie within the `zero velocity surface' (i.e. within a distance of 1060 kpc) of the barycentre, and 
for which we were able to find mention in the literature of at least one associated GC and a stellar velocity dispersion for the host galaxy. 
Given the ongoing discovery of faint satellites and their GCs \citep[e.g.][]{2016ApJ...824L..14C,2017ApJ...837...54C}, the list is probably far from complete but summarises the current situation. The number of {\it reported} GCs associated with each galaxy may go up or down over time. For example, some GCs may be missed as the search area does not reach out to a significant fraction of the virial radius (which is around a hundred times the galaxy effective radius), or they may lie below the detection limit of the imaging survey (although such faint, low-mass GCs may only have a small effect on the total GC system mass). Alternatively, the number of GCs may be overestimated if they are incorrectly identified foreground stars, nuclear clusters, background galaxies or young massive clusters within the host galaxy. Overall, the census of Local Group GCs is probably close to complete, with a slight bias towards being an underestimate of the true number. However, a final complication is that many of the Local Group galaxies are satellites that have suffered tidal stripping \citep[e.g.][]{1997AJ....113..634I,2006MNRAS.367..387R}. This can physically lower the GC content, the current dynamical halo mass, and the stellar mass, moving galaxies off of the GC-halo mass relation of isolated galaxies (for a discussion of this with regards the stellar-mass halo-mass relation of satellites, see e.g. \citealt{2018arXiv180707093R}).

In Tables 1 and 2 we list only objects for which we have some confidence that they are bona fide old-aged GCs. 
The mass of each GC system is calculated from the individual GC luminosities and assuming $M/L_V = 1.88$ \citep{2005ApJS..161..304M}, as also used by \citep{2010MNRAS.406.1967G}. 
Table 1 includes the host galaxy type,  luminosity ($M_V$), stellar mass ($M_*$), HI gas mass ($\MHI$), projected half-light radius ($\Reff$) and stellar velocity dispersion ($\sigma$)  
from the compilation of \citet{2012AJ....144....4M} or its online update (dated Sept. 2015). 
To calculate the stellar mass we adopt $M/L_V = 1$ as used by McConnachie. This is appropriate for a very metal-poor stellar population of mean age 5 Gyr. An older mean age of 10 Gyr would imply higher galaxy stellar masses by a factor of two (i.e. $M/L_V \sim 2$; \citealt{2005MNRAS.362..799M}). Although the low mass galaxies we study here are mostly very metal-poor, they do reveal a diversity of star formation histories with some revealing star formation to recent times. In a detailed study of 40 Local Group dwarfs, \citet{2014ApJ...789..147W} found those with $M_{\ast} > 10^5\,\Msun$ (as studied in this work) to have, on average, formed $\sim$70\% of their stellar mass within the last 10 Gyr. Thus, the use of $M/L_V = 1$ for our Local Group sample is a reasonable approximation, on average, but may systematically differ by factor of two for individual galaxies.
The velocity dispersion is based on a number of observed stars in the central regions of each galaxy and so can be regarded as a central velocity dispersion \citep{2012AJ....144....4M}.

\subsection{Obtaining halo masses for pressure-supported galaxies}

To obtain $M_{200}$ from $\sigma$, we first calculate the dynamical mass within (approximately) $\Reff$ using the mass estimator from \citet{2009ApJ...704.1274W}: 

\begin{equation}
\Mdyn = 580.0 \left(\frac{\Reff}{{\rm pc}}\right) \left(\frac{\sigma}{{\rm km\,s}^{-1}}\right)^2
\end{equation}

To obtain $\Mtwo$ from $\Mdyn$ we first subtract off the contribution from the stars: $\Mdyndm = \Mdyn - M_*/2$ (where the factor of two follows from the definition of the half light radius). We then correct for the adiabatic contraction of the dark matter halo due to the addition of the stars by adiabatically expanding the half light radius. For this, we follow \citet{1986ApJ...301...27B}, assuming circular orbits and angular momentum conservation. This gives:

\begin{equation}
\Reff' = \frac{M_*/2 +\Mdyndm}{\Mdyndm}\Reff
\label{eqn:adiexp}
\end{equation} 
For most galaxies in our sample, both the stellar mass and adiabatic contraction corrections are negligible, but we find that these are important for NGC\,147, NGC\,185 and NGC\,205 since these galaxies all have $M_{\ast}$ contribute significantly to the inner gravitational potential \citep[e.g.][]{2010ApJ...711..361G}. We then transform $\Mdyndm$ into halo mass ($\Mtwo$) using the coreNFW profile from \citet{Read+16}. This asymptotes to  the NFW profile at large radii, but allows us to account for the possibility of an inner dark matter core (i.e. a density profile with a logarithmic slope of 0) that can form due to `heating' from bursty star formation that counteracts the adiabatic contraction \citep[e.g.][]{2005MNRAS.356..107R}, or through exotic dark matter models or cosmology \citep[e.g.][]{2000PhRvL..84.3760S}. When transforming $\Mdyndm$ into $\Mtwo$, we assume a fixed core size parameter of $r_{\rm c} = 1.75\,R_{\rm e}'$ (see \citealt{Read+16} for details). To account for observational errors on $\sigma$, $R_{\rm e}'$ and $M_*$, we perform 5000 Monte Carlo draws of $\sigma$, $R_{\rm e}'$ and $M_*$ from their uncertainties, assuming Gaussian errors on $\sigma$ and $R_{\rm e}'$ and a flat prior on $M_*$ with an error of $\sigma_{M_*} = 0.5M_*$. For each draw, we then estimate $M_{\rm dyn,DM}(R_{\rm e}')$ and $M_{200}$, marginalising over $c_{200}$ and the inner slope of the density profile inside $r_{\rm c}$, assuming a flat prior. For the marginalisation over $c_{200}$, we assume that galaxies lie on the $M_{200}-c_{200}$ relation in $\Lambda$CDM, with a scatter in log-space of 0.1\,dex \citep{2014MNRAS.441.3359D}. The halo masses for each galaxy are listed in Table 1 (a machine-readable copy of Table 1 with errors on $\log\,\Mtwo$ is available as supplemental material).\vspace{4mm}

\subsection{Obtaining halo masses for rotationally-supported galaxies} 

Where available, we obtain $v_{\rm max}$ from rotation curves, using the peak value at the point where the rotation curve becomes flat. For NGC~6822, WLM, DDO52 and DDO210 (Aquarius) we take these values from the recent re-analysis of the Little Things data in \citet{2017MNRAS.467.2019R} and \citet{2017MNRAS.466.4159I}. For NGC~784 and UGC~1281, 
we use data from \citet{2014AstBu..69....1M}. For IC1959 and NGC~1311, we use data from \citet{1996ApJS..107...97M}. For NGC~247, we use data from \citet{1990AJ....100..641C}. For KK246, we use data from \citet{2011AJ....141..204K}. For the LMC we use the rotation curve traced by young stars from \citet{2016ApJ...832L..23V}, and for the SMC we use the HI rotation curve of \citet{2004ApJ...604..176S}. For the remainder of the dwarfs, the $v_{\rm max}$ are {\it derived from a line width profile, not a rotation curve} \citep{2014A&A...570A..13M}. This can cause significant scatter in the derived $M_{200}$ \citep[e.g.][]{2016MNRAS.455.3841B,2017MNRAS.470.1542S,2017ApJ...850...97B}. We discuss this further in Section 3. Note that several dwarfs in our sample show irregular rotation curves, indicative of them being out of equilibrium \citep[e.g.][]{2016MNRAS.462.3628R}. This includes Pegasus \citep{2016MNRAS.462.3628R} 
which we discuss further in Section 3.

We obtain $M_{200}$ from the inclination-corrected $v_{\rm max}$ in the following way. Following 
\citet{2000astro.ph..5323S}, 
we may write the maximum circular velocity for an NFW profile as:
 
\begin{equation} 
v_{\rm max} = 0.465 \frac{\sqrt{G M_{200}/{r_{200}}}}{\sqrt{c_{200}^{-1} \ln(1.0+c_{200})-(1.0+c_{200})^{-1.0}}}
\end{equation}
where $c_{200}$ is the concentration parameter and $r_{200}$ is the virial radius: 

\begin{equation}
r_{200} = \left(\frac{3}{4} \frac{M_{200}}{200 \pi \rho_{\rm crit}}\right)^{1/3}
\end{equation}
where $\rho_{\rm crit}$ is the critical density of the Universe today, i.e 136.05 $\Msun\,{\rm kpc}^{-3}$. 

We interpolate an array of $v_{\rm max}(M_{200})$ which we then use to solve numerically for $M_{200}(v_{\rm max})$. To account for errors in $v_{\rm max}$, we perform 5000 Monte Carlo draws of $v_{\rm max}$ from its error distribution, assuming Gaussian uncertainties. For each draw, we solve for $M_{200}$ while marginalising over $c_{200}$, as above\footnote{Note that when calculating $M_{200}$ from $v_{\rm max}$, we do not need to worry about the possible presence of a central dark matter core since we assume that the core size is smaller than the radius at which the rotation curve becomes flat \citep{2017MNRAS.467.2019R}. Nor do we need to worry about any correction due to the stellar mass or gas mass because these become negligible at radii where the rotation curve is flat.}. The halo masses derived in this way are listed in Table 2 (a machine-readable version, with errors on log $M_{200}$ is available as supplemental material).

In Table 3 we compare our halo masses for the Local Group dIrr galaxies with those from the literature, i.e. \citet{2017MNRAS.467.2019R} and \citet{2015AJ....149..180O}. Both of these studies fit dark matter profiles to the full HI data from the Little Things survey (a core-NFW profile for Read et al. and a NFW profile for Oh et al.). 
Our values show good agreement with both studies within our quoted uncertainties. 

\subsection{Systematic errors in the derived halo masses}

Although we estimate errors in the halo mass, as described above, there are a number of additional systematic uncertainties that are difficult to quantify. For example, the halo masses based on dynamical mass (Section 2.1) assume that the stellar system is in pressure-supported dynamical equilibrium  with very little rotational contribution and the orbits are nearly isotropic  \citep{2009AJ....137.3100W}. For nearby  early-type dwarfs moving in the tidal field of the Milky Way or Andromeda (e.g. Eridanus II and And XXV), the assumption of dynamical equilibrium may not be valid. 
An additional uncertainty comes from the application of a single mass-to-light ratio for the stars, i.e. 
$M/L_V$ = 1. As discussed earlier, this ratio could be up to twice that assumed but its effect on the halo mass is only significant for NGC 147, 185 and 205. 
In such cases 
the halo mass would be systematically overestimated.

For our late-type galaxies, 
the halo masses for most galaxies (see Section 2.2) come from integrated HI measurements over the entire galaxy rather than a resolved rotation curve. Their $v_{\rm max}$ values can be affected by asymmetries, non-circular motions, HI bubbles, incorrect inclination measurements and insufficient radial coverage so that the flat part of the rotation curve is not reached \citep[e.g.][]{2016MNRAS.462.3628R,2016MNRAS.455.3841B,2017ApJ...850...97B,2017arXiv170607478O}. We also assume the contribution from random motions to be small. 
As the HI velocities are measured at much larger radii than stellar velocity dispersions, the contribution from the baryonic mass is small and so the halo masses are not subject to a significant stellar mass correction. 
Inclination errors cause the fitted inclination to be systematically biased low, leading to an artificially low $M_{200}$ -- a problem that is exacerbated by HI bubbles that form due to stellar feedback \citep{2016MNRAS.462.3628R}. Neglecting pressure support also underestimates the mass. 
The sign of all of the above effects is to cause us to systematically {\it underestimate} the halo masses for late-type galaxies.

\subsection{Tidal Stripping}

Tidal stripping in the potential of a larger galaxy is a physical effect that will tend to lower the derived  $M_{200}$ \citep[e.g.][]{2006MNRAS.367..387R}. 
In a simulation of tides affecting Local Group dwarfs in a Milky Way like potential, \citet{2008ApJ...673..226P} found that the central velocity dispersion decreased monotonically with the amount of stripped material. The simulated galaxies initially expanded but then became more compact as the stripping removed more mass. Overall tidally stripped dwarfs have a systematically lower velocity dispersion for their size  \citep{2018MNRAS.478.3879S}. 
Thus after significant halo mass loss, the inferred dynamical mass is also reduced.

Tidal stripping tends to remove the least bound material first. This implies that after removing the extended dark matter halo, 
stripping will remove the 
GCs and finally the galaxy's stars \citep[e.g.][]{2006MNRAS.366..429R,2006MNRAS.367..387R}. Thus the relative depletion of the halo, GC system and stellar component will depend on the extent of the tidal stripping -- in mild events, the mass in stars and GCs may be relatively unaffected (and see also \citealt{2018arXiv180707093R} for a discussion of this).

\begin{table*}
\caption{Rotationally-supported dwarf galaxies with globular clusters and HI rotation measurements. }
\label{tab:georgiev}
\begin{tabular}{lccccccccc}
\hline
Name     & Type & $M_V$    & $M_*$  & $\MHI$ & log $M_{200}$ & $\Ngc$  & $\Mgc$ & $v_{\rm max}$\\
         & &  [mag] & [10$^7\,\Msun$] & [10$^{7}$ M$_{\odot}$] &  [$\Msun$] &        & [10$^4$ $\Msun$] &  [km/s] \\
\hline
LMC & Irr & -18.36 & 162.7 & 44.1 & 11.16 & 16 & 330 & 91.7 (18.8)\\
SMC & Irr & -16.82 & 36.9 & 51.2 & 10.56 & 1 & 32 & 60.0 (5.0)\\
NGC 6822 & dIrr & -15.2 & 10.3 & 13 & 10.47 & 8 & 126 & 56 (2.2)\\
WLM & dIrr & -14.2 & 4.1 & 0.61 & 9.95 & 1 & 50.4  & 39 (3.3) \\
Pegasus & dIrr & -12.2 & 0.65 & 0.59 & 8.50 & 1 & 11.5 & 13.8 (5.0)\\
Aquarius & dIrr & -10.6 & 0.15 & 0.41 & 8.82 & 1 & 1.46 & 17.8 (9.5) \\
\hline
DDO52 & Im & -14.98 & 6.38 & 19.9 & 10.32 & 2& 17.6& 50.7 (13.4)\\
ESO059-01 & IBsm & -14.60& 5.08& 8.26& 10.45 & 1& 143.9 & 55.2 (1.9)\\
ESO121-20 & Im & -13.64& 0.98& 11.4& 9.49 & 1& 4.04 & 28.1 (1.0)\\
ESO137-18 & Ssc & -17.21& 68.0& 34.1& 10.67 & 7& 111.2 & 64.5 (2.0)\\
ESO154-023 & SBm & -16.38& 19.9& 81.7& 10.43 & 3& 3.52& 54.4 (1.9)\\
ESO223-09 & IB & -16.47& 22.6& 63.8& 10.50 & 8& 307.8& 57.7 (2.7)\\
ESO269-58 & I0 & -15.78& 30.1& 2.31& 9.60 & 8& 76.9 & 29.8 (2.0)\\
ESO274-01 & Scd & -17.47& 151.1& 20.1& 10.86 & 10& 130.6 & 74 (2.5)\\
ESO381-20 & IBsm & -14.80& 5.40& 15.7& 9.91 &1& 2.28 & 37.8 (1.5)\\
ESO384-016 & dS0 & -13.72& 1.80& 0.50& 7.77 & 2& 5.83& 8.4 (0.9)\\
IC1959 & SBsm & -15.99& 14.1 & 18.7& 10.51 & 7& 172.3 & 57.4 (0.9)\\
KK16 & Irr & -12.38& 0.52& 0.69& 7.99 & 1& 2.08& 9.8 (0.3)\\
KK17 & Irr & -10.57& 0.10& 0.55& 8.16 & 1& 1.09 & 11.0 (0.5)\\
KK246 & Irr & -13.77& 0.97 & 11.9& 10.07 & 2& 41.0 & 42.0 (2.0)\\
KKH77 & Irr & -14.58& 3.98& 4.67& 9.48 & 2& 27.11 & 27.8 (1.1) \\
NGC1311 & SBm & -15.76& 11.7& 8.74& 9.89 & 5& 68.3 & 37.2 (0.4)\\
NGC247 & SBd &-18.76& 232.5 & 137.1& 11.39 & 25& 398.2 & 108 (2.0)\\
NGC4163 & Irr & -14.21& 1.92& 1.42& 8.87 & 2& 111.5 & 18.1 (0.8)\\
NGC4605 & SBd & -18.41& 172.9& 25.8&  10.58& 22& 350.4 & 60.8 (2.0)\\
NGC5237 & Irr? &-15.45& 15.5& 3.10& 9.87 & 3& 53.2 & 36.7 (1.5)\\
NGC784 & SBdm & -16.87& 84.6 & 28.9& 10.66 & 6& 40.7 & 47.5 (1.0)\\
UGC1281 & Sdm & -15.30& 12.6 & 16.4& 10.47 & 2& 31.9 & 56.4 (1.4)\\
UGC3755 & Im & -15.50& 9.28& 13.9& 8.79 & 9& 239.3 & 17.1 (0.6)\\
UGC4115 & Im & -15.12& 8.52& 20.9& 10.00 & 5& 29.9 & 40.2 (1.1)\\
UGC685 & Im & -14.35 & 3.22 & 5.88 & 10.03 & 5& 136.9 & 41.0 (1.2)\\
UGC8760 & IBm & -13.16 & 1.08 & 1.86 & 8.45 & 1 & 1.33  & 13.4 (0.3)\\
UGCA86 & Im & -16.13 & 16.5 & 48.2 & 10.59 & 11 & 1148.9  & 60.9 (2.3)\\
UGCA92 & Im & -14.71 & 4.48 & 7.62 & 9.68& 2 & 46.1  & 32.1 (1.3)\\
NGC1427A & IBsm & -18.50 & 199.4 & 176.6 & 10.16 & 38 & 605.3 & 44.9 (1.6)\\
\hline
\end{tabular}
\\\vspace{0.1cm}
Notes: The upper half of the table are Irregular galaxies in the Local Group and the lower half are isolated late-type dwarf galaxies. Galaxy type, absolute magnitude ($M_V$), stellar mass ($M_{\ast}$) and HI mass ($M_{HI}$) come from McConnachie (2012) for Local Group dwarfs and from Georgiev et al. (2010) for isolated dwarfs (and the LMC/SMC). 
The halo masses ($M_{200}$) are derived from $v_{\rm max}$ (see text for details). 
The number of GCs ($N_{\rm GC}$) and GC system mass ($M_{\rm GC}$) for isolated dwarfs come from Georgiev et al. (2010). The online version of this table includes errors on the halo mass.
\end{table*}


\section{Notes on Individual Local Group Dwarf Galaxies with GC systems}

Next, we briefly discuss the GC systems for each Local Group dwarf galaxy listed in Tables 1 and 2. 

\subsection{Sagittarius}
The Sgr dwarf galaxy and its GC system have been disrupted and merged with the Milky Way galaxy. Its original pre-merger properties are highly uncertain. 
Here we use studies that have attempted to reassemble the original Sgr galaxy and its GC system. Rather than use the basic stellar properties from \citet{2012AJ....144....4M} we take the total luminosity for Sgr from \citet{2010ApJ...712..516N}, who included the stellar streams. We take their upper limit to the  total luminosity of $L_V = 1.32 \times 10^8\,\Lsun$ and 
assume $M/L_V = 1$ (as for other Local Group galaxies)  to obtain a total stellar mass of 13.2 $\times 10^7 ~\Msun$. \citet{2017MNRAS.464..794G} have modelled Sgr plus its stream and derived a total halo mass of at least $M_{200} = 6 \times 10^{10} ~\Msun$. Here we use this halo mass rather than one based on scaling from the dynamical mass, since Sgr has definitely lost significant mass due to tidal stripping \citep[e.g.][]{2001MNRAS.323..529H,2018arXiv180707093R}. 
If the possible nucleus of the Sgr galaxy (M54) is included as a bona fide GC, then the current number of GCs likely associated with the original Sgr dwarf is 8 according to \citet{2010ApJ...718.1128L}. They are
Terzan 7 ($M_V = -5.01$), Terzan 8 ($M_V = -5.07$), Arp\,2 ($M_V = -5.29$), Pal\,12 ($M_V = -4.47$), NGC 6715 (M54, $M_V = -9.98$), Whiting\,1 ($M_V = -2.46$), NGC 5634 ($M_V = -7.69$), NGC 5053 ($M_V = -6.76$). These 8 give a total GC system mass of $1.92 \times 10^6\, \Msun$. We note that the recent proper motion study of \citet{2018arXiv180401994S} indicates that NGC 5053 is unlikely to belong to Sgr but the GC NGC 2419 ($M_V = -9.42$) does belong. In this case the GC system mass is $2.78 \times 10^6 \Msun$, which we adopt in Table 1. The resulting GC system mass to halo mass ratio for the `reassembled' Sgr dwarf is $4.6\times10^{-5}$, and hence quite consistent with the value found for higher mass galaxies.


\subsection{Fornax}
We combine the luminosities of Fornax's 5 GCs (i.e. $M_V = -5.23, -7.30, -8.19, -7.23, -7.38$) 
and a mass-to-light ratio of $M/L_V$ = 1.88 to derive a total GC system mass of $7.3\times10^5\,\Msun$.  
We note that \citet{2003MNRAS.340..175M} estimate the total mass of the Fornax dSph GCs to be $9\times10^5$ $\Msun$ while \citet{2012A&A...546A..53L} derive dynamical masses of GCs Fornax 3,4 and 5 to be $9.8\times10^5\,\Msun$. These are comparable to our estimate here.
Recently, \citet{2018arXiv180500484E} 
derive a halo mass of log $M_{200} = 9.42$ assuming a cored dark matter profile which is similar to our value in Table 1. 
However, we note that the pre-infall halo mass of Fornax was likely larger than estimated here (i.e. log $M_{200}$ = 9.4--10; \citealt{2014ApJ...782L..39A,2018arXiv180707093R}), indicative of tidal stripping and/or shocking (as is the case for the Sgr dwarf). However, we note that this scenario is  challenged by Fornax's apparently benign orbit \citep[e.g.][]{2015MNRAS.454.2401B}.

\subsection{Eridanus II}
\citet{2016ApJ...824L..14C} measured the properties of 
a faint GC ($M_V$ = --3.5) near the centre of the very low luminosity ($M_V$ = --7.1) galaxy Eridanus II.
This single GC has a corresponding mass of $4.0\times10^3$ M$_{\odot}$. The GC has a CMD that is similar to its host galaxy, i.e. dominated by an old ($\sim$10 Gyr) and metal-poor ([Fe/H] $\sim$ --2.5) stellar population. 
\citet{2017ApJ...838....8L} confirmed the low metallicity finding [Fe/H] = --2.38 and measured a velocity dispersion of 6.9 km/s. 
\citet{2016ApJ...824L..14C} also managed to measure the half-light radius of the galaxy ($\Reff\sim$ 280\,pc) and place a tight upper limit on the HI gas content of $2.8\times10^3$ $\Msun$. This suggests that either Eridanus II may interacted with the Milky Way in the past or it is a true `fossil' of reionisation \citep[e.g.][]{2011ApJ...741...17B}. Any dynamical interaction, however, must have occurred long ago given its current distance from the Milky Way of $366\pm 17$\,kpc \citep{2016ApJ...824L..14C}. Thus, we consider it unlikely that Eridanus II is significantly tidally distorted today.



\subsection{NGC 205}
NGC 205 is thought to have around a dozen GCs \citep{2000AJ....119..727B} with \citet{2000A&A...358..471F} suggesting a total GC system of 11 $\pm$ 6. Magnitudes are given for 9 GCs in \citet{1988ApJ...334..159D}. Following da Costa \& Mould, we exclude Hubble III (its radial velocity suggests an association with M31) and Hubble V (its age is less than 2 Gyr) but do include the GC M31C-55. Here we combine the masses of those 7 GCs and assume that the other 4 have a similar mean mass. This process gives us a total mass for the GC system associated with NGC 205 to be $1.41\times10^6$ $\Msun$. NGC 205 appears to be currently interacting with M31 \citep[e.g.][]{2008ApJ...683..722H} which may lead to a large systematic error in our estimate of $M_{200}$ for this galaxy \citep[e.g.][]{2006MNRAS.367..387R}. The uncertainty on our derived $M_{200}$ for NGC 205 is further compounded by its dense stellar mass. This leads to large uncertainties on the extrapolation of $M_{200}$ from the measured mass interior to the half light radius (see Section 2).


\subsection{NGC 147 and NGC 185}
Several new GCs have been discovered by the PAndAS survey bringing the total number of GCs to 10 associated with NGC 147 and 8 associated with NGC 185. The details of these GCs including their luminosities are given by \citet{2013MNRAS.435.3654V}. The two galaxies are likely tidally interacting with one another, with NGC 147 exhibiting pronounced isophotal twists \citep{2014MNRAS.445.3862C}. As with NGC 205, this tidal distortion, and the significant central concentration of stars in NGC 147 and NGC 185, leads to large uncertainties on our derived $M_{200}$ for these dwarfs (see Section 2).

\subsection{And I and XXV}
\citet{2017PASA...34...39C} find a single GC with $M_V = -3.4$ (or equivalently 
a mass of $3.68\times10^3$ $\Msun$) associated with And I. And I itself is dominated by stars older than 8\,Gyr \citep{2014ApJ...789..147W}. For And XXV, 
\citet{2016ApJ...829...26C} note a possible GC associated with $M_V = -4.9$ (or a mass of $1.46\times10^4$ $\Msun$).  
We note that And XXV has a particularly large size for its small velocity dispersion so it may be undergoing a tidal interaction with M31 \citep{2014ApJ...783....7C} and, therefore, not be in dynamical equilibrium. 
Indeed based on the measured dynamical mass, we derive a halo mass of less than 10$^8$ $\Msun$, which is the limit for atomic cooling in the early Universe \citep[e.g.][]{2006MNRAS.371..885R}. Thus our halo mass for And XXV should be considered highly uncertain and certainly a lower bound on its pre-infall halo mass.

\subsection{The SMC and LMC}
Following \citet{2010MNRAS.406.1967G}, we include the single old GC (NGC 121) in the SMC, and the 16 old GCs in the LMC, to give the total mass of the GC system in each galaxy. We note that a number of 
Milky Way GCs have been suggested to once belong to the Magellanic Clouds, e.g. Rup 106 to the SMC \citet{1995AJ....110.1664F}. As such, our estimates of the GC system mass in the Clouds are likely to be lower bounds.

\subsection{NGC 6822 (DDO209)}
\citet{2015MNRAS.452..320V} present new imaging and spectroscopy of the original GC Hubble VII and 7 more recently discovered (compact and extended) GCs, finding them to be older than 9\,Gyr. Using the luminosities given in their table 4, and $M/L_V = 1.88$, we derive a total GC system mass of $1.26\times10^6$ $\Msun$. 
Based on the motions of the GCs, \citet{2015MNRAS.452..320V} find the galaxy to be dark matter dominated. 

\subsection{WLM (DDO221)}
The WLM galaxy has been known for some time to host a single GC with a luminosity of 
$M_V$ = --8.74 \cite{2006AJ....131.1426S}. This corresponds to a mass of $5.04 \times 10^5$ $\Msun$. Virial masses in the literature for WLM include \citet{2017MNRAS.467.2019R}  log $M_{200}$ = 9.92, \citet{2017A&A...605A..55A} log $M_{200}$ = 10.07 assuming a \citet{2014MNRAS.437..415D} DM profile and 10.39 for a coreNFW profile \citep{Read+16}, and \citet{2015AJ....149..180O} log $M_{200}$ = 10.09.
Our value of the halo mass 
is within the range of these values and their quoted uncertainties. 

\subsection{Pegasus (DDO216)}
Recently, \citet{2017ApJ...837...54C} reported a single luminous GC with $M_V = -7.14$ or mass of $1.15\times10^5\,\Msun$ associated with the Pegasus galaxy. As noted in \citet{2017MNRAS.467.2019R}, Pegasus has a radially limited and highly irregular rotation curve, and so is unlikely to be dynamical equilibrium, at least in its central parts where the HI data are available. For this reason, \citet{2017MNRAS.467.2019R} classify the galaxy as a `rogue' and exclude it from further analysis. We include Pegasus in Table 2 but note that its $M_{200}$ should be treated with caution; it is almost certainly a lower bound on its true $M_{200}$.

\subsection{Aquarius (DDO210)}
\citet{1993AJ....105..894G} note the existence of a possible GC associated with the Aquarius galaxy. Here we include it in Table 2, and its GC with a luminosity of 
$M_V$ = --4.9 or mass of $1.46 \times10^4$ $\Msun$. In calculating the halo mass for the late-type galaxies we have ignored any contribution from pressure support. Aquarius has perhaps the highest contribution from pressure-support in our sample of late-type galaxies, with a stellar velocity dispersion of 7.9 km/s. Nevertheless with $v_{\rm max}/\sigma = 2.25$, this indicates that it is still rotation dominated. Indeed, our halo mass of log $M_{200} = 8.87$, that ignores the contribution from pressure support, is consistent with both \citet{2017MNRAS.467.2019R} (log $M_{200}$ = 8.83) and \citet{2015AJ....149..180O} (log $M_{200}$ = 8.80) within their quoted uncertainties.

\section{Isolated dwarf galaxies with GC systems}

Dwarf galaxies in the Local Group are influenced by, and in some cases radically transformed by, the Milky Way and Andromeda galaxies. 
As well as tidal stripping (which may remove dark matter, GCs and stars), other effects are present. Ram-pressure stripping can remove a dwarfs' ISM, causing its star formation to cease \citep[e.g.][]{2013MNRAS.433.2749G}; dwarfs can also be morphologically transformed by the tidal field of their host galaxy \citep[e.g.][]{2001ApJ...559..754M}.  \citet{2017MNRAS.467.2019R} concluded that, for these reasons, it is much easier to test the expectations of $\Lambda$CDM at low masses in isolated dwarfs than using nearby satellites.

\citet{2010MNRAS.406.1967G} compiled a list of nearby dwarf galaxies located in low-density environments with known GC systems from HST/ACS imaging. They tend to be late-type, gas-rich dwarf galaxies. 
The HST images provided accurate distances (from the tip of the red giant branch) and were of sufficiently large field-of-view to cover the entire expected GC system. Of the 55 dwarf galaxies studied, 38 revealed the presence of old GCs; 17 contained no identifiable old GCs. 
In order to avoid galaxies with large inclination corrections to their rotation velocity, we further restrict the sample to those with $i > 20^{\circ}$ \citep[e.g.][]{2016MNRAS.462.3628R}. 

In the lower part of Table 2 we include the nearby isolated dwarf galaxies of \citet{2010MNRAS.406.1967G} that meet the above criteria.
We note that although NGC 247 is located in the Sculptor group, this group is actually a loose filament of length $\sim$5 Mpc along the line-of-sight and so it is still located in a relatively low density environment.
The columns are similar to those of Table 1. 
The stellar masses for the isolated dwarfs are taken directly from \citet{2010MNRAS.406.1967G} who used colour-based $M/L_V$ ratios with a typical value of $\sim 0.75$ to convert $V$-band luminosities into stellar masses (with uncertainty of around $\pm$0.1 dex). 
We also list HI masses from \citet{2010MNRAS.406.1967G}. We note that the HI gas mass is similar to, and sometimes exceeds, the stellar mass of the galaxy; in these cases the total baryonic mass is much greater than the stellar mass alone. As through-out this work, GC system masses in Table 2 are calculated from their individual luminosities assuming $M/L_V$ = 1.88. 
Table 2 also lists the maximum rotation velocity of the HI gas ($v_{\rm max}$).
Halo masses for the isolated dwarfs are derived using the method outlined in Section 2.2 for rotationally-supported galaxies. We find no strong difference in the halo masses of galaxies with resolved rotation curves compared to those with integrated HI line widths, but see \citet{2016MNRAS.455.3841B,2017MNRAS.470.1542S,2017ApJ...850...97B} and the discussion on this in Section 2.2.




\section{The stellar mass--GC system mass relation}

Before examining the GC system mass--halo mass relation, we present relations with stellar mass. 
The relationship between the number of GCs per galaxy luminosity or stellar mass (often referred to as $S_N$ and $T_N$ respectively) is well known to have a U-shaped relation with galaxy luminosity \citep{2005ApJ...635L.137F,2008ApJ...681..197P,2010MNRAS.406.1967G}, with dwarf galaxies having an increasing fraction of their stellar mass in GCs with decreasing luminosity. This implies that either dwarf galaxies with GCs are more efficient at forming and retaining those GCs and/or they are less efficient at forming field stars. 

\begin{figure}
	\includegraphics[width=7.2cm, angle=-90]{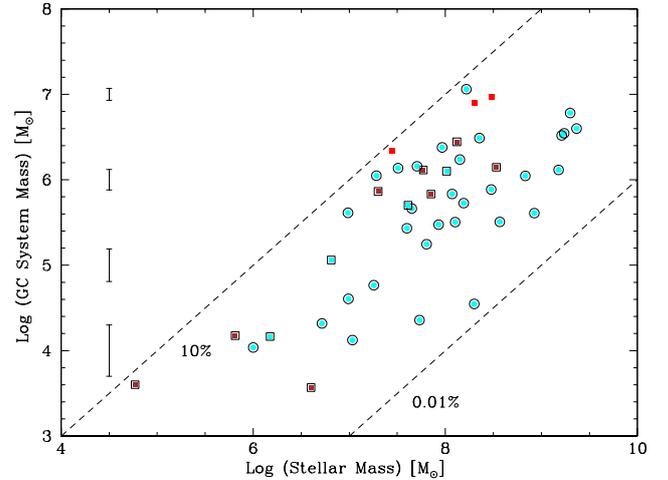}
    \caption{Stellar mass-GC system mass relation. 
  Brown symbols are dE and dSph galaxies from Table 1 and cyan symbols are late-type dwarfs from Table 2. Local Group galaxies are shown by squares and isolated galaxies by circles.
  The red squares represent ultra diffuse galaxies.  Error bars on the left hand side show representative Poisson errors in GC system mass. Dashed lines represent GC system to galaxy stellar mass ratios of 10\% and 0.01\%. 
}
    \label{fig:mstar}
\end{figure}

In Fig.~\ref{fig:mstar}, we show the GC system mass--stellar mass relation for our sample of dwarf galaxies from Tables 1 and 2 (with different symbols used). 
We estimate errors on the stellar masses to be $\pm$0.1 dex for late-type dwarfs and up to $\pm$0.3 dex for early-type dwarfs. 
For the GC system masses we assume the errors on their total luminosity to be small and the use of a single mass-to-light ratio to be reasonable (see \citealt{2009AJ....138..547S}). However, when dealing with small number statistics (e.g. sometimes a single GC) Poisson-like uncertainties may dominate. To estimate the uncertainty associated with small number statistics, we assume that GC systems with a total log $\Mgc$ = 4 has 1 GC, log $\Mgc$ = 5 has 1 GC, log $\Mgc$ = 6 has 10 GCs, log $\Mgc$ = 7 has 50 GCs on average. 
So, for example, a galaxy with 10 observed GCs may have an uncertainty of $\pm \sqrt{10}$. Assuming a similar mean GC mass, this corresponds to error bars of $\pm$0.30, 0.30, 0.12 and 0.06 dex in GC system mass. 

The data in Fig.~\ref{fig:mstar} show a large range of GC system mass to galaxy stellar mass ratios of 0.01\% to 10\%, with a typical ratio of a few tenths of a percent. An example of the range in host galaxy stellar mass for a given GC system mass is nicely illustrated by comparing Eridanus II with And I. Both galaxies are dominated by old metal-poor stars, and have a similar mass in GCs of $\sim$4 $\times$ 10$^3$ $\Msun$ but are almost 70$\times$ different in stellar mass, resulting in GC-to-stellar mass ratios of $\sim$7\% and 0.1\%.
For both galaxies there is no detected HI gas, so their total baryonic mass (i.e. stellar plus gas mass) is identical to their stellar mass. 

For comparison, the Milky Way's GC system contains 2--3\% of the  {\it halo} stellar mass, but only $\sim$0.1\% of  the {\it total} stellar mass. 
If only metal-poor stars are considered, then the GCs in some dwarf galaxies, e.g. WLM ($M_V = -14.2$) and 
IKN ($M_V = -11.5$), contain up to 30\% of the metal-poor (old) stellar content of their host galaxy \citep{2014ApJ...797...15L}. Recently, \citet{2017A&A...606A..85L} showed that, within an individual (giant) galaxy, the number of GCs is higher at lower metallicities. They argued that this was the outcome of a higher survival rate. Alternatively, dwarfs with high GC-to-stellar mass ratios may have formed most of their stars and GCs at early times, while others continued to form stars until recent times, resulting in low GC-to-stellar mass ratios. 
However, the large variation in the GC-to-stellar mass ratio for Eridanus II and And I (which are both dominated by old stars) suggests that star formation histories may not be the primary or sole driver of these variations. We also remind the reader that an increasing fraction of low mass galaxies, in the mass regime shown in Fig.~\ref{fig:mstar}, do not possess any GCs.
Further work is needed to fully understood the physical processes that give rise to the large variation in the GC system to galaxy stellar mass ratios in dwarf galaxies. 

We also include in Fig.~\ref{fig:mstar} the location of three ultra diffuse galaxies (UDGs) with measured GC systems and estimated halo masses. All three have stellar masses indicative of dwarf galaxies. The first is VCC1287 in the Virgo cluster. It has 22 $\pm$ 8 GCs and total GC system mass of around 2.2 $\pm$ 0.8 x 10$^6$ $\Msun$ \citep{2016ApJ...819L..20B}. The galaxy stellar mass is 
$2.8 \times 10^7 \Msun$. 
The second UDG is DF2 \citep{2018Natur.555..629V} 
has recently been the subject of debate over its distance. Here we adopt the larger distance of $\sim$19 Mpc 
\citep{2018arXiv180706025V}
rather than the smaller distance of 13 Mpc advocated by \citet{2018arXiv180610141T}.
Its GC system is a little odd in that it includes 11 objects with $M_V< -8.6$, i.e. more luminous than most GCs and approaching the luminosity regime of ultra compact dwarfs (UCDs). If we accept the 11 luminous compact objects, which have measured radial velocities, as bona fide GCs, then their total mass is $\sim 8 \times 10^6 \,\Msun$ (including lower mass GCs, the total system would rise to around 15 GCs). The galaxy stellar mass is $2 \times 10^8\, \Msun$. The third UDG is DF44 in the Coma cluster 
\citep{2016ApJ...828L...6V}.
It has a stellar mass of $3 \times 10^8\, \Msun$ 
and 74 $\pm{18}$ GCs 
\citep{2017ApJ...844L..11V}, 
which corresponds to a GC system mass of approximately $7.4 \times 10^6\, \Msun$. 

These three UDGs all have very high GC system mass to galaxy stellar mass ratios approaching 10\%. So although they have stellar masses similar to the dwarf galaxies considered in this work,  whatever processes have shaped these three UDGs, it has resulted in them having a large fraction of their baryonic mass in the form of GCs. This may be a result of stellar and dark matter `heating' combined with early star formation quenching \citep{2017MNRAS.466L...1D,2018MNRAS.478..906C,2018arXiv180707093R}. We note that many other UDGs have little or no GCs \citep{2018MNRAS.475.4235A}. 


\section{The stellar mass--halo mass relation (SMHR)}

There have been a number of recent studies of the stellar-mass to halo mass relation (SMHR) at low mass in the literature. These fall into three main categories: (1) using abundance matching to determine the SMHR \citep[e.g.][]{2010ApJ...717..379B,2010ApJ...710..903M,2014ApJ...784L..14B,2017MNRAS.464.3108G}; (2) measuring the SMHR for a sample of galaxies for which $M_{200}$ can be estimated \citep[e.g.][]{2015MNRAS.447..298H,2017MNRAS.466.1648K,2017MNRAS.467.2019R}; (3) modelling the SMHR using numerical simulations \citep[e.g.][]{2015MNRAS.448.2941S,2015MNRAS.454.2981C,2017arXiv170506286M}. 
and (4) semi-analytical modelling 
\citep{2017MNRAS.470..651R}.

\begin{figure}
	\includegraphics[width=7.2cm, angle=-90]{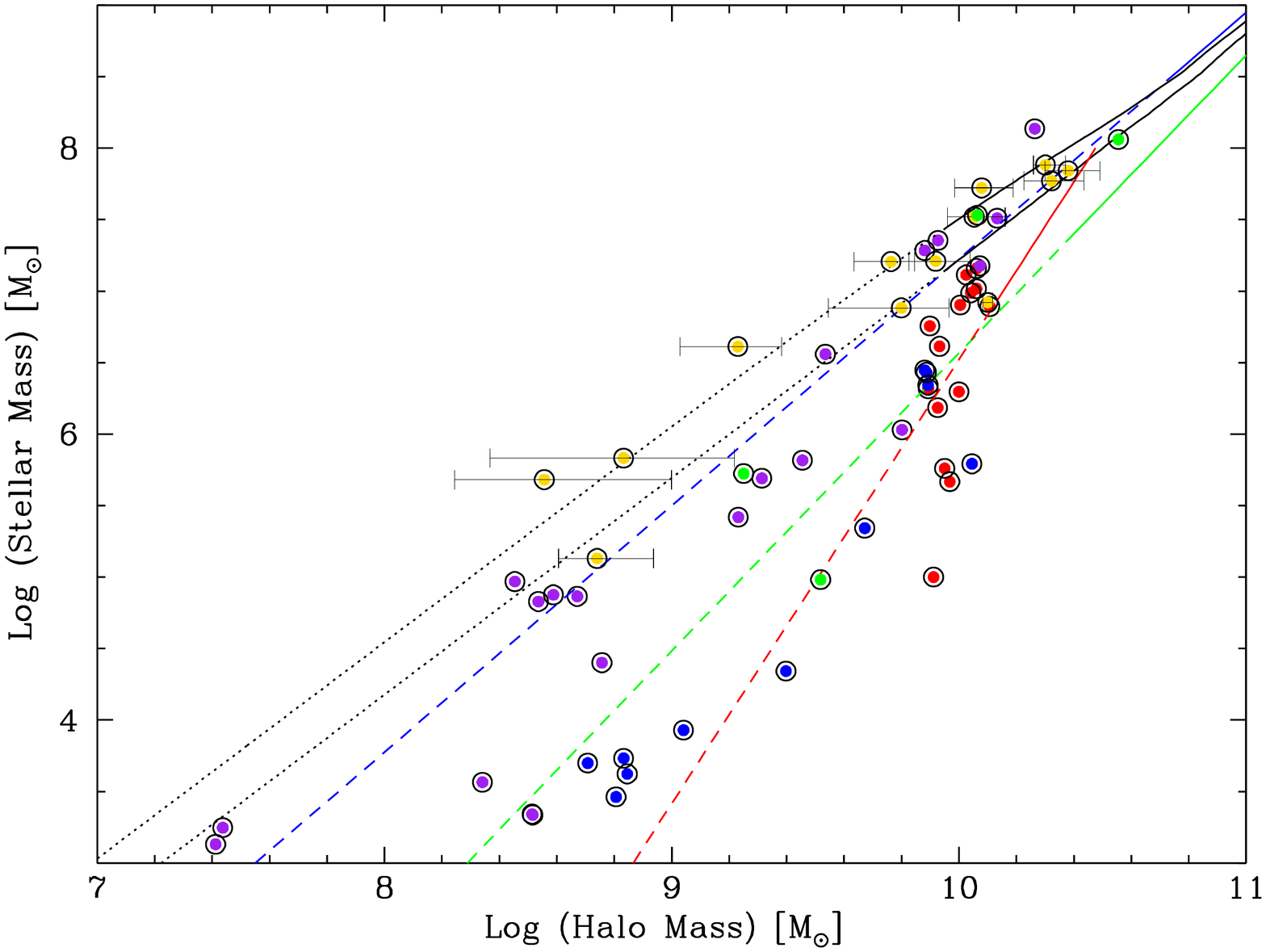}
    \caption{A compilation of stellar mass-halo mass relations taken from the literature. The filled circles show simulated dwarf galaxies 
    (purple = \citet{2017arXiv170506286M}; 
    red = \citet{2017MNRAS.471.3547F}; green = \citet{2014ApJ...792...99S}; blue = \citet{2015MNRAS.453.1305W}) and observed dwarf galaxies (gold = \citet{2017MNRAS.467.2019R}). Solid lines show predicted relations (green = \citet{2017MNRAS.470..651R}; red = \citet{2014ApJ...784L..14B}; blue = \citet{2015MNRAS.447..298H}, with extrapolations marked as dashed lines. The dotted lines show the upper and lower bounds of the extrapolated \citet{2010ApJ...717..379B} relation. 
    }
    \label{fig:smhr}
\end{figure}

In Fig.~\ref{fig:smhr} we show a selection of these studies, with relations extrapolated to lower masses.
Over the stellar mass range $10^7 < M_*/{\rm M}_\odot < 10^8$ \citet{2017MNRAS.467.2019R} find an excellent agreement between $M_{200}$ derived from abundance matching of \citet{2010ApJ...717..379B} and $M_{200}$ derived from the HI rotation curves of a sample of nearby isolated dwarfs with excellent quality data (compare the gold data points in Fig.~\ref{fig:smhr} with the dotted lines). This is consistent with the extrapolation to low mass of the similar relation (blue line) derived from weak lensing measurements by \citet{2015MNRAS.447..298H}. 
This good agreement continues down to $M_* \sim 10^5$\,M$_\odot$, however at these low stellar masses, the abundance matching relation becomes an extrapolation, while there are only a handful of galaxies with HI data that populate the plot. Thus, the relationship between stellar mass and halo mass below $M_* \sim 10^7$\,M$_\odot$ remains uncertain. 

Apparently at odds with the recent \citet{2017MNRAS.467.2019R} study, however, are two other abundance matching relations in the literature. \citet{2014ApJ...784L..14B} find a much steeper relation (red line), as do \citet{2017MNRAS.470..651R} (green line). The former study uses Milky Way satellites to reach a much lower stellar mass than is possible using the SDSS stellar mass function. However, this means that \citet{2014ApJ...784L..14B} abundance matching primarily uses {\it group satellite galaxies}. Using a stellar mass function measured from nearby groups \citep{2005RSPTA.363.2693R}, \citet{2017MNRAS.467.2019R} show that using groups does indeed lead to a much steeper SMHR than using SDSS field galaxies, explaining the difference in the SMHR of \citet{2014ApJ...784L..14B} and that of \citet{2010ApJ...717..379B} that is much shallower. The \citet{2017MNRAS.470..651R} SMHR is complete down to $M_* \sim 10^8$\,M$_\odot$ and so is an extrapolation over much of the stellar mass and halo mass range probed by our study here.

Finally, we compare the SMHR relations in Figure 2 with the latest numerical simulation results. The FIRE dwarf galaxy simulations \citep{2017MNRAS.471.3547F} cover the range 10$^5$ to 10$^7$ $\Msun$ in stellar mass, while the simulations of \citet{2014ApJ...792...99S} and \citet{2015MNRAS.453.1305W} provide a small number of additional dwarf galaxy models which probe to higher and lower stellar masses, respectively, than the \citet{2017MNRAS.471.3547F} simulations. While these simulations are consistent with one another, the \citet{2017arXiv170506286M} simulations\footnote{\citet{2017arXiv170506286M} quotes values for $M_{100}$. We convert these to $M_{200}$ using the relations in \citet{2010arXiv1005.0411C}.}, that probe a similar range in stellar mass, produce significantly more stellar mass for a given halo mass in much better agreement with the isolated dwarf sample of \citet{2017MNRAS.467.2019R} and the abundance matching of 
\citet{2010ApJ...717..379B}.

As pointed out recently by \citet{2018MNRAS.476.3124C}, most of these simulations scatter below the observed SMHR for isolated gas rich dwarfs \citep{2017MNRAS.467.2019R}, and the SMHR from abundance matching with field galaxies \citep{2010ApJ...717..379B}. Understanding the origin of this discrepancy remains a key outstanding challenge for galaxy formation models.

\begin{table}
\caption{Comparison of log halo mass estimates}
\label{tab:compare}
\begin{tabular}{lcccc}
\hline
Name & This work & R17 & O15 & P16\\
\hline 

NGC 6822 & 10.47$\pm{0.10}$ & 10.30$^{+0.04}_{-0.07}$ & -- & --\\
WLM & 9.95$\pm{0.14}$ & 9.92$^{+0.09}_{-0.12}$ & 10.09 & --\\
Pegasus & 8.50$\pm{0.55}$ & 8.83$^{+0.23}_{-0.21}$ & 9.02 & --\\
Aquarius & 8.82$\pm{0.80}$ & 8.83$^{+0.46}_{-0.38}$ & 8.80 &  --\\
\hline
NGC 247 & 11.39$\pm{0.12}$ & -- & -- & 11.10\\
\hline
\end{tabular}
\\

Notes: The table lists the halo mass from this work compared to literature works that fit the HI data to a dark matter profile. R17 = 
\citet{2017MNRAS.467.2019R}, 
O15 = \citet{2015AJ....149..180O} 
and \citet{2016PhDT.......110P}.

\end{table}

\begin{figure*}
	\includegraphics[width=14cm, angle=-90]{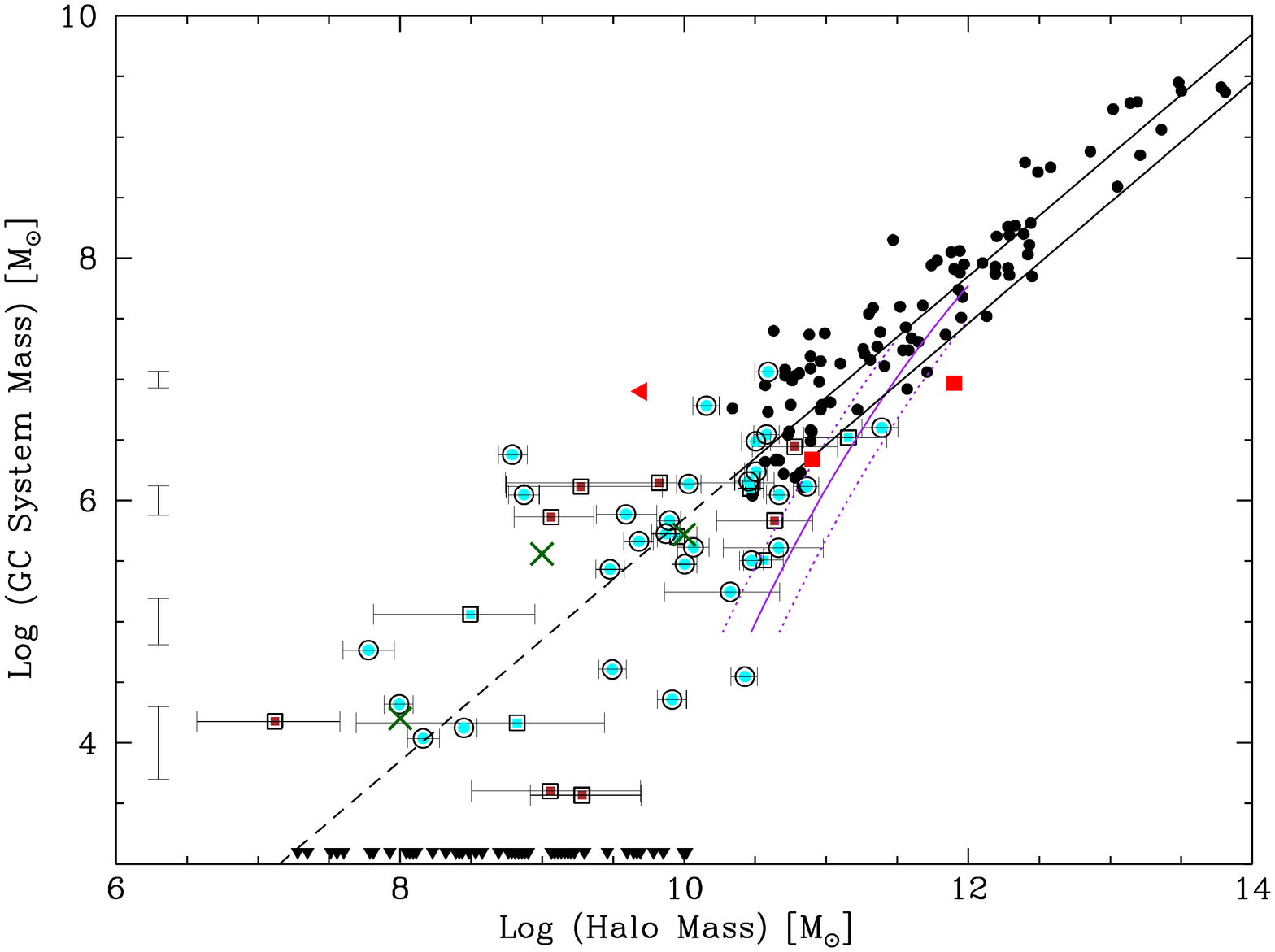}
    \caption{GC system mass--halo mass. Brown symbols are dE and dSph galaxies from Table 1 and cyan symbols are late-type dwarfs from Table 2 (with 1$\sigma$ error bars). Local Group galaxies are shown by squares and isolated galaxies by circles. The green crosses represent the median value in 3 halo mass bins centered at 8,9 and 10 in log halo mass. 
    Small black symbols are galaxies from \citet{2009MNRAS.392L...1S}. 
    Error bars on the left hand side show representative Poisson errors in GC system mass. 
    The red squares indicate the UDG galaxies DF44 and VCC1287 and the red triangle the UDG DF2 with an upper limit to its halo mass (see text for details).
    Solid triangles on the lower edge of the plot represent Local Group galaxies without known GC systems (see text for details). 
    The dashed and dotted lines are the relations of \citet{2009MNRAS.392L...1S} and \citet{2017ApJ...836...67H} 
    extrapolated to lower masses. The purple lines show the approximate predicted relation, and scatter, of  the semi-analytic model of \citet{2018arXiv180503652E} at low masses.
	}
    \label{fig:halo}
\end{figure*}

\section{Globular cluster system mass--halo mass relation}

To recap, in Table 1 we list the Local Group dE and dSph galaxies with known GC systems. The census of GCs in the Local Group is largely complete. However, the galaxies are satellites and may be subject to tidal effects. The halo mass of the host galaxy is calculated  from the measured dynamical mass within the half light radius (see section 2.1).   
Table 2 lists the Local Group dIrr galaxies and isolated late-type galaxies from \citet{2010MNRAS.406.1967G} with both GCs and HI rotation kinematics. The isolated galaxies, which dominate our sample, are largely free of tidal effects but they are not statistically complete in terms of their GC census. 
Halo masses are calculated 
from their observed inclination-corrected maximum rotation velocity (see section 2.2). For all galaxies, we calculate the total mass of their GC system from the individual reported GCs.

The key result from this paper is shown in  Fig.~\ref{fig:halo}. Here 
we show the GC system mass--halo mass relation for the dwarf galaxies listed in Tables 1 and 2, along with the median value of GC system mass in three halo mass bins of width 1 dex.
We show the 
linear relations of \citet{2009MNRAS.392L...1S} and \citet{2017ApJ...836...67H} 
extrapolated to lower masses. These relations  
(which use different mean GC masses in their derivation) are based on photometry of and on the assumption of selecting predominately old GCs.  
We also show the recent prediction from the semi-analytical model of \citet{2018arXiv180503652E}. We include the estimated uncertainty in halo mass for each galaxy and show representative errors for the GC system mass assuming a Poisson-like uncertainty in GC counting statistics (as described in Section 5). 

{\it The GC systems of very low mass galaxies are consistent with a linear extrapolation of the Spitler \& Forbes relation from higher mass, albeit with significant scatter.} This scatter is larger than our estimated 1$\sigma$ error bars. Thus, it could be driven by real cosmic variation in the GC-to-halo mass ratio. However, we caution that the systematic errors on $M_{200}$ are substantial and likely bias our halo estimates low (see Section 2). We see no clear difference between Local Group and isolated galaxies.

Our low mass galaxies do not follow the prediction of the fiducial model of \citet{2018arXiv180503652E}. 
They used a semi-analytical model based on dark matter merger trees to predict the GC system mass--halo mass relation.
Their fiducial model matches the observed one at high masses but starts to deviate for log $M_{200} \le 12.0$ and is significantly different to our trend at their lowest halo mass of log $M_{200}\sim 10.5$. They suggest that the linear relation seen at high mass is the simple result of a self-similar merger history of halos with old GCs. By contrast, lower mass galaxies have fewer mergers and the predicted relation curves downwards with less mass in GCs for a given halo mass.
They assumed a minimum GC mass of 10$^5$ $\Msun$ and did not include any mass-dependent disruption or tidal stripping of GCs in their model. A smaller minimum mass and/or the inclusion of disruption 
would tend to further lower the predicted mass of their GC systems. 
They also suggest 
that halos with log $M_{200}<11$ will not contain  metal-rich GCs. 

Interestingly, \citet{2018arXiv180503652E} also present a `random' model (their figure 3) in which there is no initial correlation of GC system and halo mass at early times, but in which they do allow growth via mergers. A large number of mergers acts to preserve the linear relation down to very low masses. This can be thought of as a consequence of the central limit theorem, which also produces a tight relation at high masses and increasing scatter towards low mass as we observe in Fig.~\ref{fig:halo}. The random model also predicts that the majority of GCs are metal-poor, again as observed in low-mass galaxies \citep{2010MNRAS.406.1967G}. 

So, although the fiducial model of \citet{2018arXiv180503652E} does not correctly reproduce the properties of the GC systems in our low mass galaxies, their random model does a much better job. This suggests that GC formation in at least some low-mass halos in the early Universe must be more efficient than assumed in their fiducial model. Indeed, \citet{2009ApJ...706L.192B} argue for just such an efficient GC formation at early times in low mass halos in order to reproduce the observed concentrated distribution of old GCs around massive galaxies like the Milky Way today (and see also \citealt{2006MNRAS.368..563M} and \citealt{2018MNRAS.tmp.1537K}). Such an increased efficiency in early star formation may also be required to bring the simulations of low mass dwarfs into agreement with the data for nearby gas-rich isolated dwarfs (see Figure 2, \citealt{2018MNRAS.476.3124C} and \citealt{2017MNRAS.467.2019R}). 

Our results suggest that at least some halos with $M_{200}< 10^{10}\,\Msun$ form GCs. However, below $M_{200}\simeq10^{10}\,\Msun$, many galaxies -- perhaps even the majority -- appear to be be devoid of GCs (see Fig.~\ref{fig:halo},  black triangles). Thus, when averaging over all galaxies, the {\it mean} GC system mass per unit of halo mass appears to fall more steeply than the GC system mass-halo mass relation of dwarfs selected to contain GCs. Such stochasticity in the GC occupancy of dwarfs may allow for a better agreement between our data and the fiducial El-Badry et al. model (if the fiducial El-Badry et al. model were adapted to allow for stochastic GC occupation). We discuss this issue in more detail in Section~\ref{sec:nogcs}. 

We also include in Fig.~\ref{fig:halo} the location of the three ultra diffuse galaxies (UDGs) presented in Section 5. 
VCC1287 has a galaxy total mass, based on the motions of its GCs, of $M_{200} = 8 \pm 4 \times 10^{10}\,\Msun$ \citep{2016ApJ...819L..20B}. This implies a GC system to total mass ratio of $5.5 \times 10^{-5}$, which is consistent with the other galaxies in this plot. 
Based on its stellar velocity dispersion, DF44 
has an estimated halo mass of $8 \times 10^{11}\, \Msun$ \citep{2016ApJ...828L...6V} and is an outlier in the plot. 
\citet{2018Natur.555..629V} infer a 90\% confidence upper limit on the halo mass of DF2, from both the stellar velocity dispersion and the motions of its GCs, of $< 1.5\times 10^8\, \Msun$. 
We note that both \citet{2018ApJ...859L...5M} and \citet{2018arXiv180404139L} have questioned the dynamical mass calculation of \citet{2018Natur.555..629V}, with Laporte et al. suggesting its halo mass could be up to $5 \times 10^9\,\Msun$.
In Fig.~\ref{fig:halo}, we show the upper limit to the halo mass  from Laporte et al. 
Although the three UDGs discussed here have stellar masses similar to those of the low mass galaxies studied in this work, their GC system masses and halo masses are more indicative of massive galaxies. 
The diversity in their GC system and halo mass properties clearly  
warrants further study.

\begin{figure}
	\includegraphics[width=7.2cm, angle=-90]{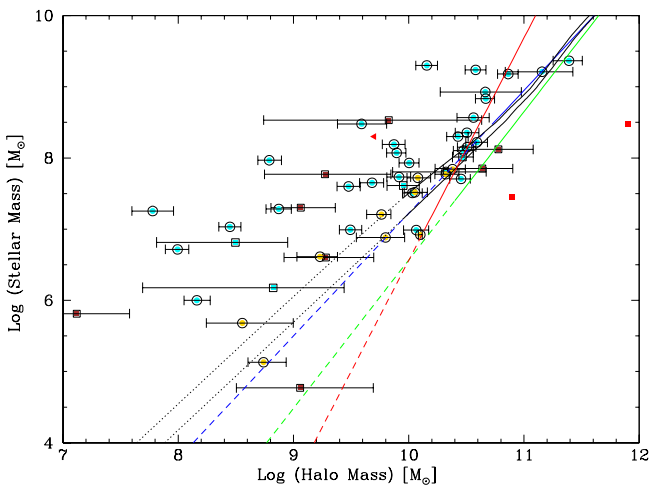}
    \caption{A compilation of stellar mass-halo mass relations taken from the literature. The filled circles show observed dwarf galaxies with GC systems; brown = Table 1 dE and dSph galaxies, cyan = Table 2 dIrr galaxies). Local Group galaxies are shown by squares, while isolated galaxies are shown by circles.
    Gold filled circles show isolated galaxies without GCs from \citet{2017MNRAS.467.2019R}.  
    Solid lines show predicted relations (green = \citet{2017MNRAS.470..651R}; red = \citet{2014ApJ...784L..14B}; blue = \citet{2015MNRAS.447..298H}, with extrapolations marked as dashed lines. The dotted lines show the upper and lower bounds of the \citet{2010ApJ...717..379B} relation. 
        The red squares indicate the UDGs DF44 and VCC1287, and the red triangle UDG DF2.  
}
    \label{fig:smhrgc}
\end{figure}

Finally, we plot in Fig.~\ref{fig:smhrgc} the SMHR for our dwarf galaxies with GC systems. 
We remind the reader that there is an uncertainty in the stellar mass of $\pm$0.3 dex for the dE/dSph galaxies and $\pm$0.1 dex for the late-type dwarfs. 
Our galaxies with GC systems have systematically lower halo masses than even the shallowest SMHR relation shown, with a slope of around unity, noting that systematic effects tend to bias our halo masses towards lower limits. 
Our data show a similar trend to the `clean sample' of relatively isolated dwarfs studied by \citet{2017MNRAS.467.2019R}.  
As our galaxies (with GC systems) and the galaxies shown from Read et al. (without GCs) are dominated by isolated galaxies, it is unlikely that tidal stripping is the main cause of the reduced halo masses. 

\section{Galaxies without globular clusters}
\label{sec:nogcs}

In this paper we have mainly considered galaxies with GC systems, however galaxies without GC systems appear to become increasingly common at the lowest masses (see Fig \ref{fig:halo}, black triangles). The lack of a GC system in some dwarf galaxies could be due to their removal by tidal stripping (i.e. an environmental effect), their destruction over cosmic time (which may include sinking to the galaxy centre via dynamical friction) or that they never formed in the first place.  We note that since \citet{2010MNRAS.406.1967G}
found several isolated galaxies lacking GC systems, 
the absence of GCs cannot be purely due to tidal stripping.

In Fig.~\ref{fig:halo}, we qualitatively represent the Local Group galaxies (taken from \citealt{2012AJ....144....4M})
that lack reported GCs, showing their halo mass distribution. The Local Group sample should be close to a complete census of GCs, however as it is not a low density environment their GC systems may have been previously depleted due to tidal stripping.
Taking a simple approach of using the stellar mass directly from \citet{2012AJ....144....4M} and their halo mass assuming an extrapolation of the SMHR relation in Fig.~\ref{fig:smhr} of \citet{2015MNRAS.447..298H},
Fig.~\ref{fig:halo} shows that an increasing fraction of low mass Local Group dwarf galaxies (with halo masses log $M_{200}$ $<$ 10) 
lack GC systems. Any study interested in the {\it mean} mass of GC systems as a function of halo mass needs to take account of galaxies with zero GC system mass. Indeed, using this qualitative approach we estimate that the fraction of mass contained in GC systems is at least ten times lower in these lower mass halos. 

In this work, we have 
investigated the {\it populated} GC system mass--halo mass relation, i.e. only galaxies with reported GC systems.
The fiducial and random models presented by  
\citet{2018arXiv180503652E} 
are also {\it populated} relations (i.e. only for galaxies with GCs) and not {\it mean} relations for
all galaxies with, and without, GC systems. Their semi-analytic model does not include stochastic effects, such as halo-to-halo variation of the baryon fraction, ISM properties and environmental effects which become increasingly important in the low mass regime. 
Based on the qualitative halo mass distribution shown in Fig.~\ref{fig:halo}, we would expect the {\it mean} GC system mass to become non-linear and decrease with decreasing halo mass in the low mass regime. In other words, low mass galaxies will have a smaller {\it mean} GC system mass to halo mass ratio than universally found for high mass galaxies.


The question of the lowest mass galaxy that can host a single GC was raised in the Introduction. 
\citet{2018arXiv180503652E} 
noted that {\it most} halos with log $M_{200} < 10.5$ (or log $M_{\ast} \le 5$) in their simulation did not form a GC. 
Our data show that many local dwarfs with $M_{200}<10.5$  
do form GCs, and they do so with a scatter about the same linear relation as those of higher mass galaxies. Only for halo masses 
log $M_{200}<10$ do Local Group galaxies without GCs become a common feature. \cite{2018arXiv180503652E}
note that their relation would continue to be linear to lower masses if GC formation was more efficient in low mass halos at early times than in their fiducial model.


\citet{2017MNRAS.472.3120B} introduced a model to explain the GC system mass vs halo mass relation. His model assumes that all halos with total mass above 10$^9$ M$_{\odot}$ at high redshift (i.e. $z\sim6$) can form a GC system whose total mass is directly proportional to the halo mass. As the total GC system mass at redshift zero is directly related to the sum of the progenitor halo masses during galaxy assembly, a linear relation between GC system mass and host galaxy halo mass is produced. This model predicts that the minimum galaxy halo mass (at $z = 0$) to host a single GC will be $\sim 10^{10}$ $\Msun$, if metal-poor GC formation ends at $z = 6$. In particular, the probability of hosting at least 1 GC is 10\% for a halo mass of $10^{10}$ M$_{\odot}$ 
but $<1$\% for a halo of mass $7\times10^9$ M$_{\odot}$. Here we find several Local Group galaxies with a halo mass less than $7\times10^{9}$ M$_{\odot}$ that host a GC system.
If metal-poor GC formation ended at a lower redshift (which is more compatible with the ages of GCs), then the minimum halo mass would be a few times larger and in better agreement with the number of Local Group dwarfs with GCs. 
\citet{2010MNRAS.406.1967G} suggested that the lower stellar mass limit for the host galaxy would be around 3.5$\times$10$^5$ $\Msun$. Since then, Eridanus II with a stellar mass of 
5.9$\times$10$^4$ $\Msun$ has been found to host a single (albeit low mass) GC.

\begin{table*}
\caption{Nearby dwarf galaxies with globular clusters but lacking kinematic data. 
}
\begin{tabular}{llccccc}
\hline
Name & Type & $M_V$ & $M_{\ast}$ & $N_{GC}$ & $M_{GC}$ & Ref \\
     & & [mag] & [10$^6$ $\Msun$] &   & [10$^4$ $\Msun$] & \\
     \hline
Phoenix & dIrr & -9.9 & 0.77 & 4 & 3.1 & M99\\
KKs3 & dE & -10.8 & 4.5 & 1 & 35  & K15\\
IKN & dSph & -11.5 & 3.4 &  5 & 80 & L14\\
ScI-dE1 & dE & -11.5  & 3.4 & 1 & 7.5 &  D09\\
\hline
\end{tabular}
\\
Notes: For KKs3 we quote the B band magnitude. 
References: M99 = Martinez-Deldago et al. (1999), 
K15 = Karachentsev et al. (2015), L14 = Larsen et al. (2014), D09 = da Costa et al. (2009). 
\end{table*}

Future observational studies interested in the {\it mean} GC system -- halo mass relation need to consider galaxies without GCs. Ideally, one requires a volume-complete census of isolated galaxies (to reduce the effects of tidal stripping on the their halos and GC systems) that is deep (to include faint GCs) and that searches a wide area around each galaxy (to include 
GCs at large galactocentric radii). Such a detailed observational study does not yet exist. 
A study of the {\it mean} relation should also calculate the halo mass based on their observed kinematics. 
As well as GC-free galaxies, such studies may also need to consider star-free galaxies (i.e. dark halos) that could dominate the population of galaxies with halo masses below $M_{200}\sim 10^9$ $\Msun$ \citep{2017MNRAS.471.3547F}.

\section{Conclusions}

In this work, we have collected a sample of early and late-type Local Group dwarf galaxies and supplemented this with a larger sample of nearby late-type dwarf galaxies in low-density environments. All of the galaxies in our sample are known to host at least one bona fide old globular cluster (GC). We estimate the total mass of each GC system based on the combined luminosity of its individual GCs and a fixed mass-to-light ratio. We estimate host galaxy halo mass from measured stellar kinematics in the case of the early-type dwarfs and from HI gas kinematics in the case of the late-type dwarfs. For galaxies where literature estimates of the halo mass are available, we find reasonable agreement with our halo masses, within our respective uncertainties.

We find that for the lowest mass galaxies, there is a wide range in the mass of GC systems relative to their host galaxy stellar mass, i.e. from about 0.01\% to over 10\% of the total stellar mass is contained in GCs. Some ultra diffuse galaxies also have relative fractions close to 10\%. 


In this work, we have used halo masses estimated from observed kinematics, to show that the mass of a GC system relative to its host galaxy halo mass is consistent, albeit with large scatter, with a simple extrapolation of the linear GC system mass--halo mass relation from higher masses. 
Thus, the near-linear relation which ranges from halo masses of 10$^{14}$ to 10$^{10}$ $\Msun$ extends down to halo masses of $\le$10$^9$ $\Msun$ without a substantial change in slope. 
The scatter has however increased considerably to around $\pm$1 dex, and appears to exceed that of our estimated 1$\sigma$ error bars (however we caution that this could be due to systematic uncertainties in our $M_{200}$ estimates). In particular, `ultra diffuse' galaxies appear to have a wide range in their GC system mass to halo mass properties.

Our observations are similar to the `random' merger model of \citet{2018arXiv180503652E} but not their fiducial semi-analytic model that includes their preferred GC formation physics.  The ratio of GC system mass to halo mass is relatively constant at $\sim$0.005\% for the halo mass range 10$^8$ $<$ $M_{200}$/$\Msun$ $<$ 10$^{14}$. (However, we note that if galaxies without GCs are included, there is a decrease of the mean GC system mass per unit halo mass below $M_{200}\simeq10^{10}\,\Msun$.) The extension of this relation to the lowest masses confirms the close scaling between GCs and galaxy dark matter halos. This may be the simple consequence of many halo mergers since their formation at early times. We find that galaxy halos with mass $M_{200}$ $\le 10^{9}$\,$\Msun$ do appear to be able to host a globular cluster.

Finally, we identify several nearby dwarf galaxies with no reported GCs, for which  a careful search is likely to be fruitful.






\section*{Acknowledgements}

We thank A. Alabi, S. Larsen, A. Romanowsky and M. Hudson for useful discussions. We also thank the referee for several useful suggestions. DAF thanks the Santander Fellowship and the ARC via DP160101608 for their financial support.
JIR would like to acknowledge support from SNF grant PP00P2\_128540/1, STFC consolidated grant ST/M000990/1 and the MERAC foundation. MG acknowledges financial support from the Royal Society (University Research Fellowship) and the
European Research Council (ERC StG-335936, CLUSTERS).



\bibliographystyle{mnras}
\bibliography{bib} 

\begin{thebibliography}{}
\makeatletter
\relax
\def\mn@urlcharsother{\let\do\@makeother \do\$\do\&\do\#\do\^\do\_\do\%\do\~}
\def\mn@doi{\begingroup\mn@urlcharsother \@ifnextchar [ {\mn@doi@}
  {\mn@doi@[]}}
\def\mn@doi@[#1]#2{\def\@tempa{#1}\ifx\@tempa\@empty \href
  {http://dx.doi.org/#2} {doi:#2}\else \href {http://dx.doi.org/#2} {#1}\fi
  \endgroup}
\def\mn@eprint#1#2{\mn@eprint@#1:#2::\@nil}
\def\mn@eprint@arXiv#1{\href {http://arxiv.org/abs/#1} {{\tt arXiv:#1}}}
\def\mn@eprint@dblp#1{\href {http://dblp.uni-trier.de/rec/bibtex/#1.xml}
  {dblp:#1}}
\def\mn@eprint@#1:#2:#3:#4\@nil{\def\@tempa {#1}\def\@tempb {#2}\def\@tempc
  {#3}\ifx \@tempc \@empty \let \@tempc \@tempb \let \@tempb \@tempa \fi \ifx
  \@tempb \@empty \def\@tempb {arXiv}\fi \@ifundefined
  {mn@eprint@\@tempb}{\@tempb:\@tempc}{\expandafter \expandafter \csname
  mn@eprint@\@tempb\endcsname \expandafter{\@tempc}}}

\bibitem[\protect\citeauthoryear{{Allaert}, {Gentile}  \& {Baes}}{{Allaert}
  et~al.}{2017}]{2017A&A...605A..55A}
{Allaert} F.,  {Gentile} G.,   {Baes} M.,  2017, \mn@doi [\aap]
  {10.1051/0004-6361/201730402}, \href
  {http://adsabs.harvard.edu/abs/2017A%26A...605A..55A} {605, A55}

\bibitem[\protect\citeauthoryear{{Amorisco}, {Zavala}  \& {de Boer}}{{Amorisco}
  et~al.}{2014}]{2014ApJ...782L..39A}
{Amorisco} N.~C.,  {Zavala} J.,   {de Boer} T.~J.~L.,  2014, \mn@doi [\apjl]
  {10.1088/2041-8205/782/2/L39}, \href
  {http://adsabs.harvard.edu/abs/2014ApJ...782L..39A} {782, L39}

\bibitem[\protect\citeauthoryear{{Amorisco}, {Monachesi}, {Agnello}  \&
  {White}}{{Amorisco} et~al.}{2018}]{2018MNRAS.475.4235A}
{Amorisco} N.~C.,  {Monachesi} A.,  {Agnello} A.,   {White} S.~D.~M.,  2018,
  \mn@doi [\mnras] {10.1093/mnras/sty116}, \href
  {http://adsabs.harvard.edu/abs/2018MNRAS.475.4235A} {475, 4235}

\bibitem[\protect\citeauthoryear{{Barmby}, {Huchra}, {Brodie}, {Forbes},
  {Schroder}  \& {Grillmair}}{{Barmby} et~al.}{2000}]{2000AJ....119..727B}
{Barmby} P.,  {Huchra} J.~P.,  {Brodie} J.~P.,  {Forbes} D.~A.,  {Schroder}
  L.~L.,   {Grillmair} C.~J.,  2000, \mn@doi [\aj] {10.1086/301213}, \href
  {http://adsabs.harvard.edu/abs/2000AJ....119..727B} {119, 727}

\bibitem[\protect\citeauthoryear{{Battaglia}, {Sollima}  \&
  {Nipoti}}{{Battaglia} et~al.}{2015}]{2015MNRAS.454.2401B}
{Battaglia} G.,  {Sollima} A.,   {Nipoti} C.,  2015, \mn@doi [\mnras]
  {10.1093/mnras/stv2096}, \href
  {http://adsabs.harvard.edu/abs/2015MNRAS.454.2401B} {454, 2401}

\bibitem[\protect\citeauthoryear{{Beasley}, {Romanowsky}, {Pota}, {Navarro},
  {Martinez Delgado}, {Neyer}  \& {Deich}}{{Beasley}
  et~al.}{2016}]{2016ApJ...819L..20B}
{Beasley} M.~A.,  {Romanowsky} A.~J.,  {Pota} V.,  {Navarro} I.~M.,  {Martinez
  Delgado} D.,  {Neyer} F.,   {Deich} A.~L.,  2016, \mn@doi [\apjl]
  {10.3847/2041-8205/819/2/L20}, \href
  {http://adsabs.harvard.edu/abs/2016ApJ...819L..20B} {819, L20}

\bibitem[\protect\citeauthoryear{{Behroozi}, {Conroy}  \&
  {Wechsler}}{{Behroozi} et~al.}{2010}]{2010ApJ...717..379B}
{Behroozi} P.~S.,  {Conroy} C.,   {Wechsler} R.~H.,  2010, \mn@doi [\apj]
  {10.1088/0004-637X/717/1/379}, \href
  {http://adsabs.harvard.edu/abs/2010ApJ...717..379B} {717, 379}

\bibitem[\protect\citeauthoryear{{Bellazzini} et~al.,}{{Bellazzini}
  et~al.}{2014}]{2014A&A...566A..44B}
{Bellazzini} M.,  et~al., 2014, \mn@doi [\aap] {10.1051/0004-6361/201423659},
  \href {http://adsabs.harvard.edu/abs/2014A%26A...566A..44B} {566, A44}

\bibitem[\protect\citeauthoryear{{Belokurov}, {Irwin}, {Koposov}, {Evans},
  {Gonzalez-Solares}, {Metcalfe}  \& {Shanks}}{{Belokurov}
  et~al.}{2014}]{2014MNRAS.441.2124B}
{Belokurov} V.,  {Irwin} M.~J.,  {Koposov} S.~E.,  {Evans} N.~W.,
  {Gonzalez-Solares} E.,  {Metcalfe} N.,   {Shanks} T.,  2014, \mn@doi [\mnras]
  {10.1093/mnras/stu626}, \href
  {http://adsabs.harvard.edu/abs/2014MNRAS.441.2124B} {441, 2124}

\bibitem[\protect\citeauthoryear{{Blumenthal}, {Faber}, {Flores}  \&
  {Primack}}{{Blumenthal} et~al.}{1986}]{1986ApJ...301...27B}
{Blumenthal} G.~R.,  {Faber} S.~M.,  {Flores} R.,   {Primack} J.~R.,  1986,
  \mn@doi [\apj] {10.1086/163867}, \href
  {http://adsabs.harvard.edu/abs/1986ApJ...301...27B} {301, 27}

\bibitem[\protect\citeauthoryear{{Boley}, {Lake}, {Read}  \&
  {Teyssier}}{{Boley} et~al.}{2009}]{2009ApJ...706L.192B}
{Boley} A.~C.,  {Lake} G.,  {Read} J.,   {Teyssier} R.,  2009, \mn@doi [\apjl]
  {10.1088/0004-637X/706/1/L192}, \href
  {http://adsabs.harvard.edu/abs/2009ApJ...706L.192B} {706, L192}

\bibitem[\protect\citeauthoryear{{Bovill} \& {Ricotti}}{{Bovill} \&
  {Ricotti}}{2011}]{2011ApJ...741...17B}
{Bovill} M.~S.,  {Ricotti} M.,  2011, \mn@doi [\apj]
  {10.1088/0004-637X/741/1/17}, \href
  {http://adsabs.harvard.edu/abs/2011ApJ...741...17B} {741, 17}

\bibitem[\protect\citeauthoryear{{Boylan-Kolchin}}{{Boylan-Kolchin}}{2017}]{2017MNRAS.472.3120B}
{Boylan-Kolchin} M.,  2017, \mn@doi [\mnras] {10.1093/mnras/stx2164}, \href
  {http://adsabs.harvard.edu/abs/2017MNRAS.472.3120B} {472, 3120}

\bibitem[\protect\citeauthoryear{{Brook} \& {Shankar}}{{Brook} \&
  {Shankar}}{2016}]{2016MNRAS.455.3841B}
{Brook} C.~B.,  {Shankar} F.,  2016, \mn@doi [\mnras] {10.1093/mnras/stv2550},
  \href {http://adsabs.harvard.edu/abs/2016MNRAS.455.3841B} {455, 3841}

\bibitem[\protect\citeauthoryear{{Brook}, {Di Cintio}, {Knebe},
  {Gottl{\"o}ber}, {Hoffman}, {Yepes}  \& {Garrison-Kimmel}}{{Brook}
  et~al.}{2014}]{2014ApJ...784L..14B}
{Brook} C.~B.,  {Di Cintio} A.,  {Knebe} A.,  {Gottl{\"o}ber} S.,  {Hoffman}
  Y.,  {Yepes} G.,   {Garrison-Kimmel} S.,  2014, \mn@doi [\apjl]
  {10.1088/2041-8205/784/1/L14}, \href
  {http://adsabs.harvard.edu/abs/2014ApJ...784L..14B} {784, L14}

\bibitem[\protect\citeauthoryear{{Brooks}, {Papastergis}, {Christensen},
  {Governato}, {Stilp}, {Quinn}  \& {Wadsley}}{{Brooks}
  et~al.}{2017}]{2017ApJ...850...97B}
{Brooks} A.~M.,  {Papastergis} E.,  {Christensen} C.~R.,  {Governato} F.,
  {Stilp} A.,  {Quinn} T.~R.,   {Wadsley} J.,  2017, \mn@doi [\apj]
  {10.3847/1538-4357/aa9576}, \href
  {http://adsabs.harvard.edu/abs/2017ApJ...850...97B} {850, 97}

\bibitem[\protect\citeauthoryear{{Caldwell}, {Strader}, {Sand}, {Willman}  \&
  {Seth}}{{Caldwell} et~al.}{2017}]{2017PASA...34...39C}
{Caldwell} N.,  {Strader} J.,  {Sand} D.~J.,  {Willman} B.,   {Seth} A.~C.,
  2017, \mn@doi [\pasa] {10.1017/pasa.2017.35}, \href
  {http://adsabs.harvard.edu/abs/2017PASA...34...39C} {34, e039}

\bibitem[\protect\citeauthoryear{{Carignan} \& {Puche}}{{Carignan} \&
  {Puche}}{1990}]{1990AJ....100..641C}
{Carignan} C.,  {Puche} D.,  1990, \mn@doi [\aj] {10.1086/115547}, \href
  {http://adsabs.harvard.edu/abs/1990AJ....100..641C} {100, 641}

\bibitem[\protect\citeauthoryear{{Chan}, {Kere{\v s}}, {O{\~n}orbe}, {Hopkins},
  {Muratov}, {Faucher-Gigu{\`e}re}  \& {Quataert}}{{Chan}
  et~al.}{2015}]{2015MNRAS.454.2981C}
{Chan} T.~K.,  {Kere{\v s}} D.,  {O{\~n}orbe} J.,  {Hopkins} P.~F.,  {Muratov}
  A.~L.,  {Faucher-Gigu{\`e}re} C.-A.,   {Quataert} E.,  2015, \mn@doi [\mnras]
  {10.1093/mnras/stv2165}, \href
  {http://adsabs.harvard.edu/abs/2015MNRAS.454.2981C} {454, 2981}

\bibitem[\protect\citeauthoryear{{Chan}, {Kere{\v s}}, {Wetzel}, {Hopkins},
  {Faucher-Gigu{\`e}re}, {El-Badry}, {Garrison-Kimmel}  \&
  {Boylan-Kolchin}}{{Chan} et~al.}{2018}]{2018MNRAS.478..906C}
{Chan} T.~K.,  {Kere{\v s}} D.,  {Wetzel} A.,  {Hopkins} P.~F.,
  {Faucher-Gigu{\`e}re} C.-A.,  {El-Badry} K.,  {Garrison-Kimmel} S.,
  {Boylan-Kolchin} M.,  2018, \mn@doi [\mnras] {10.1093/mnras/sty1153}, \href
  {http://adsabs.harvard.edu/abs/2018MNRAS.478..906C} {478, 906}

\bibitem[\protect\citeauthoryear{{Coe}}{{Coe}}{2010}]{2010arXiv1005.0411C}
{Coe} D.,  2010, preprint, \href
  {http://adsabs.harvard.edu/abs/2010arXiv1005.0411C} {} (\mn@eprint {arXiv}
  {1005.0411})

\bibitem[\protect\citeauthoryear{{Cole} et~al.,}{{Cole}
  et~al.}{2017}]{2017ApJ...837...54C}
{Cole} A.~A.,  et~al., 2017, \mn@doi [\apj] {10.3847/1538-4357/aa5df6}, \href
  {http://adsabs.harvard.edu/abs/2017ApJ...837...54C} {837, 54}

\bibitem[\protect\citeauthoryear{{Collins} et~al.,}{{Collins}
  et~al.}{2014}]{2014ApJ...783....7C}
{Collins} M.~L.~M.,  et~al., 2014, \mn@doi [\apj] {10.1088/0004-637X/783/1/7},
  \href {http://adsabs.harvard.edu/abs/2014ApJ...783....7C} {783, 7}

\bibitem[\protect\citeauthoryear{{Contenta} et~al.,}{{Contenta}
  et~al.}{2018}]{2018MNRAS.476.3124C}
{Contenta} F.,  et~al., 2018, \mn@doi [\mnras] {10.1093/mnras/sty424}, \href
  {http://adsabs.harvard.edu/abs/2018MNRAS.476.3124C} {476, 3124}

\bibitem[\protect\citeauthoryear{{Crnojevi{\'c}} et~al.,}{{Crnojevi{\'c}}
  et~al.}{2014}]{2014MNRAS.445.3862C}
{Crnojevi{\'c}} D.,  et~al., 2014, \mn@doi [\mnras] {10.1093/mnras/stu2003},
  \href {http://adsabs.harvard.edu/abs/2014MNRAS.445.3862C} {445, 3862}

\bibitem[\protect\citeauthoryear{{Crnojevi{\'c}}, {Sand}, {Zaritsky},
  {Spekkens}, {Willman}  \& {Hargis}}{{Crnojevi{\'c}}
  et~al.}{2016}]{2016ApJ...824L..14C}
{Crnojevi{\'c}} D.,  {Sand} D.~J.,  {Zaritsky} D.,  {Spekkens} K.,  {Willman}
  B.,   {Hargis} J.~R.,  2016, \mn@doi [\apjl] {10.3847/2041-8205/824/1/L14},
  \href {http://adsabs.harvard.edu/abs/2016ApJ...824L..14C} {824, L14}

\bibitem[\protect\citeauthoryear{{Cusano} et~al.,}{{Cusano}
  et~al.}{2016}]{2016ApJ...829...26C}
{Cusano} F.,  et~al., 2016, \mn@doi [\apj] {10.3847/0004-637X/829/1/26}, \href
  {http://adsabs.harvard.edu/abs/2016ApJ...829...26C} {829, 26}

\bibitem[\protect\citeauthoryear{{Da Costa} \& {Mould}}{{Da Costa} \&
  {Mould}}{1988}]{1988ApJ...334..159D}
{Da Costa} G.~S.,  {Mould} J.~R.,  1988, \mn@doi [\apj] {10.1086/166826}, \href
  {http://adsabs.harvard.edu/abs/1988ApJ...334..159D} {334, 159}

\bibitem[\protect\citeauthoryear{{Dekel} \& {Birnboim}}{{Dekel} \&
  {Birnboim}}{2006}]{2006MNRAS.368....2D}
{Dekel} A.,  {Birnboim} Y.,  2006, \mn@doi [\mnras]
  {10.1111/j.1365-2966.2006.10145.x}, \href
  {http://adsabs.harvard.edu/abs/2006MNRAS.368....2D} {368, 2}

\bibitem[\protect\citeauthoryear{{Di Cintio}, {Brook}, {Macci{\`o}}, {Stinson},
  {Knebe}, {Dutton}  \& {Wadsley}}{{Di Cintio}
  et~al.}{2014}]{2014MNRAS.437..415D}
{Di Cintio} A.,  {Brook} C.~B.,  {Macci{\`o}} A.~V.,  {Stinson} G.~S.,  {Knebe}
  A.,  {Dutton} A.~A.,   {Wadsley} J.,  2014, \mn@doi [\mnras]
  {10.1093/mnras/stt1891}, \href
  {http://adsabs.harvard.edu/abs/2014MNRAS.437..415D} {437, 415}

\bibitem[\protect\citeauthoryear{{Di Cintio}, {Brook}, {Dutton}, {Macci{\`o}},
  {Obreja}  \& {Dekel}}{{Di Cintio} et~al.}{2017}]{2017MNRAS.466L...1D}
{Di Cintio} A.,  {Brook} C.~B.,  {Dutton} A.~A.,  {Macci{\`o}} A.~V.,  {Obreja}
  A.,   {Dekel} A.,  2017, \mn@doi [\mnras] {10.1093/mnrasl/slw210}, \href
  {http://adsabs.harvard.edu/abs/2017MNRAS.466L...1D} {466, L1}

\bibitem[\protect\citeauthoryear{{Dutton} \& {Macci{\`o}}}{{Dutton} \&
  {Macci{\`o}}}{2014}]{2014MNRAS.441.3359D}
{Dutton} A.~A.,  {Macci{\`o}} A.~V.,  2014, \mn@doi [\mnras]
  {10.1093/mnras/stu742}, \href
  {http://adsabs.harvard.edu/abs/2014MNRAS.441.3359D} {441, 3359}

\bibitem[\protect\citeauthoryear{{El-Badry}, {Quataert}, {Weisz}, {Choksi}  \&
  {Boylan-Kolchin}}{{El-Badry} et~al.}{2018}]{2018arXiv180503652E}
{El-Badry} K.,  {Quataert} E.,  {Weisz} D.~R.,  {Choksi} N.,   {Boylan-Kolchin}
  M.,  2018, preprint, \href
  {http://adsabs.harvard.edu/abs/2018arXiv180503652E} {} (\mn@eprint {arXiv}
  {1805.03652})

\bibitem[\protect\citeauthoryear{{Errani}, {Pe{\~n}arrubia}  \&
  {Walker}}{{Errani} et~al.}{2018}]{2018arXiv180500484E}
{Errani} R.,  {Pe{\~n}arrubia} J.,   {Walker} M.~G.,  2018, preprint, \href
  {http://adsabs.harvard.edu/abs/2018arXiv180500484E} {} (\mn@eprint {arXiv}
  {1805.00484})

\bibitem[\protect\citeauthoryear{{Fitts} et~al.,}{{Fitts}
  et~al.}{2017}]{2017MNRAS.471.3547F}
{Fitts} A.,  et~al., 2017, \mn@doi [\mnras] {10.1093/mnras/stx1757}, \href
  {http://adsabs.harvard.edu/abs/2017MNRAS.471.3547F} {471, 3547}

\bibitem[\protect\citeauthoryear{{Forbes}}{{Forbes}}{2005}]{2005ApJ...635L.137F}
{Forbes} D.~A.,  2005, \mn@doi [\apjl] {10.1086/499563}, \href
  {http://adsabs.harvard.edu/abs/2005ApJ...635L.137F} {635, L137}

\bibitem[\protect\citeauthoryear{{Forbes}, {Masters}, {Minniti}  \&
  {Barmby}}{{Forbes} et~al.}{2000}]{2000A&A...358..471F}
{Forbes} D.~A.,  {Masters} K.~L.,  {Minniti} D.,   {Barmby} P.,  2000, \aap,
  \href {http://adsabs.harvard.edu/abs/2000A%26A...358..471F} {358, 471}

\bibitem[\protect\citeauthoryear{{Fusi Pecci}, {Bellazzini}, {Cacciari}  \&
  {Ferraro}}{{Fusi Pecci} et~al.}{1995}]{1995AJ....110.1664F}
{Fusi Pecci} F.,  {Bellazzini} M.,  {Cacciari} C.,   {Ferraro} F.~R.,  1995,
  \mn@doi [\aj] {10.1086/117639}, \href
  {http://adsabs.harvard.edu/abs/1995AJ....110.1664F} {110, 1664}

\bibitem[\protect\citeauthoryear{{Garrison-Kimmel}, {Bullock}, {Boylan-Kolchin}
   \& {Bardwell}}{{Garrison-Kimmel} et~al.}{2017}]{2017MNRAS.464.3108G}
{Garrison-Kimmel} S.,  {Bullock} J.~S.,  {Boylan-Kolchin} M.,   {Bardwell} E.,
  2017, \mn@doi [\mnras] {10.1093/mnras/stw2564}, \href
  {http://adsabs.harvard.edu/abs/2017MNRAS.464.3108G} {464, 3108}

\bibitem[\protect\citeauthoryear{{Gatto}, {Fraternali}, {Read}, {Marinacci},
  {Lux}  \& {Walch}}{{Gatto} et~al.}{2013}]{2013MNRAS.433.2749G}
{Gatto} A.,  {Fraternali} F.,  {Read} J.~I.,  {Marinacci} F.,  {Lux} H.,
  {Walch} S.,  2013, \mn@doi [\mnras] {10.1093/mnras/stt896}, \href
  {http://adsabs.harvard.edu/abs/2013MNRAS.433.2749G} {433, 2749}

\bibitem[\protect\citeauthoryear{{Geha}, {van der Marel}, {Guhathakurta},
  {Gilbert}, {Kalirai}  \& {Kirby}}{{Geha} et~al.}{2010}]{2010ApJ...711..361G}
{Geha} M.,  {van der Marel} R.~P.,  {Guhathakurta} P.,  {Gilbert} K.~M.,
  {Kalirai} J.,   {Kirby} E.~N.,  2010, \mn@doi [\apj]
  {10.1088/0004-637X/711/1/361}, \href
  {http://adsabs.harvard.edu/abs/2010ApJ...711..361G} {711, 361}

\bibitem[\protect\citeauthoryear{{Georgiev}, {Puzia}, {Goudfrooij}  \&
  {Hilker}}{{Georgiev} et~al.}{2010}]{2010MNRAS.406.1967G}
{Georgiev} I.~Y.,  {Puzia} T.~H.,  {Goudfrooij} P.,   {Hilker} M.,  2010,
  \mn@doi [\mnras] {10.1111/j.1365-2966.2010.16802.x}, \href
  {http://adsabs.harvard.edu/abs/2010MNRAS.406.1967G} {406, 1967}

\bibitem[\protect\citeauthoryear{{Gibbons}, {Belokurov}  \& {Evans}}{{Gibbons}
  et~al.}{2017}]{2017MNRAS.464..794G}
{Gibbons} S.~L.~J.,  {Belokurov} V.,   {Evans} N.~W.,  2017, \mn@doi [\mnras]
  {10.1093/mnras/stw2328}, \href
  {http://adsabs.harvard.edu/abs/2017MNRAS.464..794G} {464, 794}

\bibitem[\protect\citeauthoryear{{Greggio}, {Marconi}, {Tosi}  \&
  {Focardi}}{{Greggio} et~al.}{1993}]{1993AJ....105..894G}
{Greggio} L.,  {Marconi} G.,  {Tosi} M.,   {Focardi} P.,  1993, \mn@doi [\aj]
  {10.1086/116481}, \href {http://adsabs.harvard.edu/abs/1993AJ....105..894G}
  {105, 894}

\bibitem[\protect\citeauthoryear{{Harris}, {Harris}  \& {Alessi}}{{Harris}
  et~al.}{2013}]{2013ApJ...772...82H}
{Harris} W.~E.,  {Harris} G.~L.~H.,   {Alessi} M.,  2013, \mn@doi [\apj]
  {10.1088/0004-637X/772/2/82}, \href
  {http://adsabs.harvard.edu/abs/2013ApJ...772...82H} {772, 82}

\bibitem[\protect\citeauthoryear{{Harris}, {Harris}  \& {Hudson}}{{Harris}
  et~al.}{2015}]{2015ApJ...806...36H}
{Harris} W.~E.,  {Harris} G.~L.,   {Hudson} M.~J.,  2015, \mn@doi [\apj]
  {10.1088/0004-637X/806/1/36}, \href
  {http://adsabs.harvard.edu/abs/2015ApJ...806...36H} {806, 36}

\bibitem[\protect\citeauthoryear{{Harris}, {Blakeslee}  \& {Harris}}{{Harris}
  et~al.}{2017}]{2017ApJ...836...67H}
{Harris} W.~E.,  {Blakeslee} J.~P.,   {Harris} G.~L.~H.,  2017, \mn@doi [\apj]
  {10.3847/1538-4357/836/1/67}, \href
  {http://adsabs.harvard.edu/abs/2017ApJ...836...67H} {836, 67}

\bibitem[\protect\citeauthoryear{{Helmi} \& {White}}{{Helmi} \&
  {White}}{2001}]{2001MNRAS.323..529H}
{Helmi} A.,  {White} S.~D.~M.,  2001, \mn@doi [\mnras]
  {10.1046/j.1365-8711.2001.04238.x}, \href
  {http://adsabs.harvard.edu/abs/2001MNRAS.323..529H} {323, 529}

\bibitem[\protect\citeauthoryear{{Howley}, {Geha}, {Guhathakurta},
  {Montgomery}, {Laughlin}  \& {Johnston}}{{Howley}
  et~al.}{2008}]{2008ApJ...683..722H}
{Howley} K.~M.,  {Geha} M.,  {Guhathakurta} P.,  {Montgomery} R.~M.,
  {Laughlin} G.,   {Johnston} K.~V.,  2008, \mn@doi [\apj] {10.1086/589632},
  \href {http://adsabs.harvard.edu/abs/2008ApJ...683..722H} {683, 722}

\bibitem[\protect\citeauthoryear{{Hudson}, {Harris}  \& {Harris}}{{Hudson}
  et~al.}{2014}]{2014ApJ...787L...5H}
{Hudson} M.~J.,  {Harris} G.~L.,   {Harris} W.~E.,  2014, \mn@doi [\apjl]
  {10.1088/2041-8205/787/1/L5}, \href
  {http://adsabs.harvard.edu/abs/2014ApJ...787L...5H} {787, L5}

\bibitem[\protect\citeauthoryear{{Hudson} et~al.,}{{Hudson}
  et~al.}{2015}]{2015MNRAS.447..298H}
{Hudson} M.~J.,  et~al., 2015, \mn@doi [\mnras] {10.1093/mnras/stu2367}, \href
  {http://adsabs.harvard.edu/abs/2015MNRAS.447..298H} {447, 298}

\bibitem[\protect\citeauthoryear{{Ibata}, {Wyse}, {Gilmore}, {Irwin}  \&
  {Suntzeff}}{{Ibata} et~al.}{1997}]{1997AJ....113..634I}
{Ibata} R.~A.,  {Wyse} R.~F.~G.,  {Gilmore} G.,  {Irwin} M.~J.,   {Suntzeff}
  N.~B.,  1997, \mn@doi [\aj] {10.1086/118283}, \href
  {http://adsabs.harvard.edu/abs/1997AJ....113..634I} {113, 634}

\bibitem[\protect\citeauthoryear{{Iorio}, {Fraternali}, {Nipoti}, {Di Teodoro},
  {Read}  \& {Battaglia}}{{Iorio} et~al.}{2017}]{2017MNRAS.466.4159I}
{Iorio} G.,  {Fraternali} F.,  {Nipoti} C.,  {Di Teodoro} E.,  {Read} J.~I.,
  {Battaglia} G.,  2017, \mn@doi [\mnras] {10.1093/mnras/stw3285}, \href
  {http://adsabs.harvard.edu/abs/2017MNRAS.466.4159I} {466, 4159}

\bibitem[\protect\citeauthoryear{{Irwin}, {Ferguson}, {Huxor}, {Tanvir},
  {Ibata}  \& {Lewis}}{{Irwin} et~al.}{2008}]{2008ApJ...676L..17I}
{Irwin} M.~J.,  {Ferguson} A.~M.~N.,  {Huxor} A.~P.,  {Tanvir} N.~R.,  {Ibata}
  R.~A.,   {Lewis} G.~F.,  2008, \mn@doi [\apjl] {10.1086/587100}, \href
  {http://adsabs.harvard.edu/abs/2008ApJ...676L..17I} {676, L17}

\bibitem[\protect\citeauthoryear{{Karachentsev}, {Kniazev}  \&
  {Sharina}}{{Karachentsev} et~al.}{2015}]{2015AN....336..707K}
{Karachentsev} I.~D.,  {Kniazev} A.~Y.,   {Sharina} M.~E.,  2015, \mn@doi
  [Astronomische Nachrichten] {10.1002/asna.201512207}, \href
  {http://adsabs.harvard.edu/abs/2015AN....336..707K} {336, 707}

\bibitem[\protect\citeauthoryear{{Katz}, {Lelli}, {McGaugh}, {Di Cintio},
  {Brook}  \& {Schombert}}{{Katz} et~al.}{2017}]{2017MNRAS.466.1648K}
{Katz} H.,  {Lelli} F.,  {McGaugh} S.~S.,  {Di Cintio} A.,  {Brook} C.~B.,
  {Schombert} J.~M.,  2017, \mn@doi [\mnras] {10.1093/mnras/stw3101}, \href
  {http://adsabs.harvard.edu/abs/2017MNRAS.466.1648K} {466, 1648}

\bibitem[\protect\citeauthoryear{{Kravtsov} \& {Gnedin}}{{Kravtsov} \&
  {Gnedin}}{2005}]{2005ApJ...623..650K}
{Kravtsov} A.~V.,  {Gnedin} O.~Y.,  2005, \mn@doi [\apj] {10.1086/428636},
  \href {http://adsabs.harvard.edu/abs/2005ApJ...623..650K} {623, 650}

\bibitem[\protect\citeauthoryear{{Kreckel}, {Peebles}, {van Gorkom}, {van de
  Weygaert}  \& {van der Hulst}}{{Kreckel} et~al.}{2011}]{2011AJ....141..204K}
{Kreckel} K.,  {Peebles} P.~J.~E.,  {van Gorkom} J.~H.,  {van de Weygaert} R.,
   {van der Hulst} J.~M.,  2011, \mn@doi [\aj] {10.1088/0004-6256/141/6/204},
  \href {http://adsabs.harvard.edu/abs/2011AJ....141..204K} {141, 204}

\bibitem[\protect\citeauthoryear{{Kruijssen}, {Pfeffer}, {Reina-Campos},
  {Crain}  \& {Bastian}}{{Kruijssen} et~al.}{2018}]{2018MNRAS.tmp.1537K}
{Kruijssen} J.~M.~D.,  {Pfeffer} J.~L.,  {Reina-Campos} M.,  {Crain} R.~A.,
  {Bastian} N.,  2018, \mn@doi [\mnras] {10.1093/mnras/sty1609}, \href
  {http://adsabs.harvard.edu/abs/2018MNRAS.tmp.1537K} {}

\bibitem[\protect\citeauthoryear{{Laevens} et~al.,}{{Laevens}
  et~al.}{2014}]{2014ApJ...786L...3L}
{Laevens} B.~P.~M.,  et~al., 2014, \mn@doi [\apjl]
  {10.1088/2041-8205/786/1/L3}, \href
  {http://adsabs.harvard.edu/abs/2014ApJ...786L...3L} {786, L3}

\bibitem[\protect\citeauthoryear{{Lamers}, {Kruijssen}, {Bastian}, {Rejkuba},
  {Hilker}  \& {Kissler-Patig}}{{Lamers} et~al.}{2017}]{2017A&A...606A..85L}
{Lamers} H.~J.~G.~L.~M.,  {Kruijssen} J.~M.~D.,  {Bastian} N.,  {Rejkuba} M.,
  {Hilker} M.,   {Kissler-Patig} M.,  2017, \mn@doi [\aap]
  {10.1051/0004-6361/201731062}, \href
  {http://adsabs.harvard.edu/abs/2017A%26A...606A..85L} {606, A85}

\bibitem[\protect\citeauthoryear{{Laporte}, {Agnello}  \& {Navarro}}{{Laporte}
  et~al.}{2018}]{2018arXiv180404139L}
{Laporte} C.~F.~P.,  {Agnello} A.,   {Navarro} J.~F.,  2018, preprint, \href
  {http://adsabs.harvard.edu/abs/2018arXiv180404139L} {} (\mn@eprint {arXiv}
  {1804.04139})

\bibitem[\protect\citeauthoryear{{Larsen}, {Brodie}  \& {Strader}}{{Larsen}
  et~al.}{2012}]{2012A&A...546A..53L}
{Larsen} S.~S.,  {Brodie} J.~P.,   {Strader} J.,  2012, \mn@doi [\aap]
  {10.1051/0004-6361/201219895}, \href
  {http://adsabs.harvard.edu/abs/2012A%26A...546A..53L} {546, A53}

\bibitem[\protect\citeauthoryear{{Larsen}, {Brodie}, {Forbes}  \&
  {Strader}}{{Larsen} et~al.}{2014a}]{2014A&A...565A..98L}
{Larsen} S.~S.,  {Brodie} J.~P.,  {Forbes} D.~A.,   {Strader} J.,  2014a,
  \mn@doi [\aap] {10.1051/0004-6361/201322672}, \href
  {http://adsabs.harvard.edu/abs/2014A%26A...565A..98L} {565, A98}

\bibitem[\protect\citeauthoryear{{Larsen}, {Brodie}, {Grundahl}  \&
  {Strader}}{{Larsen} et~al.}{2014b}]{2014ApJ...797...15L}
{Larsen} S.~S.,  {Brodie} J.~P.,  {Grundahl} F.,   {Strader} J.,  2014b,
  \mn@doi [\apj] {10.1088/0004-637X/797/1/15}, \href
  {http://adsabs.harvard.edu/abs/2014ApJ...797...15L} {797, 15}

\bibitem[\protect\citeauthoryear{{Law} \& {Majewski}}{{Law} \&
  {Majewski}}{2010}]{2010ApJ...718.1128L}
{Law} D.~R.,  {Majewski} S.~R.,  2010, \mn@doi [\apj]
  {10.1088/0004-637X/718/2/1128}, \href
  {http://adsabs.harvard.edu/abs/2010ApJ...718.1128L} {718, 1128}

\bibitem[\protect\citeauthoryear{{Li} et~al.,}{{Li}
  et~al.}{2017}]{2017ApJ...838....8L}
{Li} T.~S.,  et~al., 2017, \mn@doi [\apj] {10.3847/1538-4357/aa6113}, \href
  {http://adsabs.harvard.edu/abs/2017ApJ...838....8L} {838, 8}

\bibitem[\protect\citeauthoryear{{Lim} \& {Lee}}{{Lim} \&
  {Lee}}{2015}]{2015ApJ...804..123L}
{Lim} S.,  {Lee} M.~G.,  2015, \mn@doi [\apj] {10.1088/0004-637X/804/2/123},
  \href {http://adsabs.harvard.edu/abs/2015ApJ...804..123L} {804, 123}

\bibitem[\protect\citeauthoryear{{Mackey} \& {Gilmore}}{{Mackey} \&
  {Gilmore}}{2003}]{2003MNRAS.340..175M}
{Mackey} A.~D.,  {Gilmore} G.~F.,  2003, \mn@doi [\mnras]
  {10.1046/j.1365-8711.2003.06275.x}, \href
  {http://adsabs.harvard.edu/abs/2003MNRAS.340..175M} {340, 175}

\bibitem[\protect\citeauthoryear{{Makarov}, {Prugniel}, {Terekhova}, {Courtois}
   \& {Vauglin}}{{Makarov} et~al.}{2014}]{2014A&A...570A..13M}
{Makarov} D.,  {Prugniel} P.,  {Terekhova} N.,  {Courtois} H.,   {Vauglin} I.,
  2014, \mn@doi [\aap] {10.1051/0004-6361/201423496}, \href
  {http://adsabs.harvard.edu/abs/2014A%26A...570A..13M} {570, A13}

\bibitem[\protect\citeauthoryear{{Maraston}}{{Maraston}}{2005}]{2005MNRAS.362..799M}
{Maraston} C.,  2005, \mn@doi [\mnras] {10.1111/j.1365-2966.2005.09270.x},
  \href {http://adsabs.harvard.edu/abs/2005MNRAS.362..799M} {362, 799}

\bibitem[\protect\citeauthoryear{{Martin}, {Collins}, {Longeard}  \&
  {Tollerud}}{{Martin} et~al.}{2018}]{2018ApJ...859L...5M}
{Martin} N.~F.,  {Collins} M.~L.~M.,  {Longeard} N.,   {Tollerud} E.,  2018,
  \mn@doi [\apjl] {10.3847/2041-8213/aac216}, \href
  {http://adsabs.harvard.edu/abs/2018ApJ...859L...5M} {859, L5}

\bibitem[\protect\citeauthoryear{{Mart{\'{\i}}nez-Delgado}, {Gallart}  \&
  {Aparicio}}{{Mart{\'{\i}}nez-Delgado} et~al.}{1999}]{1999AJ....118..862M}
{Mart{\'{\i}}nez-Delgado} D.,  {Gallart} C.,   {Aparicio} A.,  1999, \mn@doi
  [\aj] {10.1086/300967}, \href
  {http://adsabs.harvard.edu/abs/1999AJ....118..862M} {118, 862}

\bibitem[\protect\citeauthoryear{{Mathewson} \& {Ford}}{{Mathewson} \&
  {Ford}}{1996}]{1996ApJS..107...97M}
{Mathewson} D.~S.,  {Ford} V.~L.,  1996, \mn@doi [\apjs] {10.1086/192356},
  \href {http://adsabs.harvard.edu/abs/1996ApJS..107...97M} {107, 97}

\bibitem[\protect\citeauthoryear{{Mayer}, {Governato}, {Colpi}, {Moore},
  {Quinn}, {Wadsley}, {Stadel}  \& {Lake}}{{Mayer}
  et~al.}{2001}]{2001ApJ...559..754M}
{Mayer} L.,  {Governato} F.,  {Colpi} M.,  {Moore} B.,  {Quinn} T.,  {Wadsley}
  J.,  {Stadel} J.,   {Lake} G.,  2001, \mn@doi [\apj] {10.1086/322356}, \href
  {http://adsabs.harvard.edu/abs/2001ApJ...559..754M} {559, 754}

\bibitem[\protect\citeauthoryear{{McConnachie}}{{McConnachie}}{2012}]{2012AJ....144....4M}
{McConnachie} A.~W.,  2012, \mn@doi [\aj] {10.1088/0004-6256/144/1/4}, \href
  {http://adsabs.harvard.edu/abs/2012AJ....144....4M} {144, 4}

\bibitem[\protect\citeauthoryear{{McLaughlin} \& {van der Marel}}{{McLaughlin}
  \& {van der Marel}}{2005}]{2005ApJS..161..304M}
{McLaughlin} D.~E.,  {van der Marel} R.~P.,  2005, \mn@doi [\apjs]
  {10.1086/497429}, \href {http://adsabs.harvard.edu/abs/2005ApJS..161..304M}
  {161, 304}

\bibitem[\protect\citeauthoryear{{Moiseev}}{{Moiseev}}{2014}]{2014AstBu..69....1M}
{Moiseev} A.~V.,  2014, \mn@doi [Astrophysical Bulletin]
  {10.1134/S1990341314010015}, \href
  {http://adsabs.harvard.edu/abs/2014AstBu..69....1M} {69, 1}

\bibitem[\protect\citeauthoryear{{Moore}, {Diemand}, {Madau}, {Zemp}  \&
  {Stadel}}{{Moore} et~al.}{2006}]{2006MNRAS.368..563M}
{Moore} B.,  {Diemand} J.,  {Madau} P.,  {Zemp} M.,   {Stadel} J.,  2006,
  \mn@doi [\mnras] {10.1111/j.1365-2966.2006.10116.x}, \href
  {http://adsabs.harvard.edu/abs/2006MNRAS.368..563M} {368, 563}

\bibitem[\protect\citeauthoryear{{Moster}, {Somerville}, {Maulbetsch}, {van den
  Bosch}, {Macci{\`o}}, {Naab}  \& {Oser}}{{Moster}
  et~al.}{2010}]{2010ApJ...710..903M}
{Moster} B.~P.,  {Somerville} R.~S.,  {Maulbetsch} C.,  {van den Bosch} F.~C.,
  {Macci{\`o}} A.~V.,  {Naab} T.,   {Oser} L.,  2010, \mn@doi [\apj]
  {10.1088/0004-637X/710/2/903}, \href
  {http://adsabs.harvard.edu/abs/2010ApJ...710..903M} {710, 903}

\bibitem[\protect\citeauthoryear{{Munshi}, {Brooks}, {Applebaum}, {Weisz},
  {Governato}  \& {Quinn}}{{Munshi} et~al.}{2017}]{2017arXiv170506286M}
{Munshi} F.,  {Brooks} A.~M.,  {Applebaum} E.,  {Weisz} D.~R.,  {Governato} F.,
    {Quinn} T.~R.,  2017, preprint, \href
  {http://adsabs.harvard.edu/abs/2017arXiv170506286M} {} (\mn@eprint {arXiv}
  {1705.06286})

\bibitem[\protect\citeauthoryear{{Niederste-Ostholt}, {Belokurov}, {Evans}  \&
  {Pe{\~n}arrubia}}{{Niederste-Ostholt} et~al.}{2010}]{2010ApJ...712..516N}
{Niederste-Ostholt} M.,  {Belokurov} V.,  {Evans} N.~W.,   {Pe{\~n}arrubia} J.,
   2010, \mn@doi [\apj] {10.1088/0004-637X/712/1/516}, \href
  {http://adsabs.harvard.edu/abs/2010ApJ...712..516N} {712, 516}

\bibitem[\protect\citeauthoryear{{Oh} et~al.,}{{Oh}
  et~al.}{2015}]{2015AJ....149..180O}
{Oh} S.-H.,  et~al., 2015, \mn@doi [\aj] {10.1088/0004-6256/149/6/180}, \href
  {http://adsabs.harvard.edu/abs/2015AJ....149..180O} {149, 180}

\bibitem[\protect\citeauthoryear{{Oman}, {Marasco}, {Navarro}, {Frenk},
  {Schaye}  \& {Ben{\'{\i}}tez-Llambay}}{{Oman}
  et~al.}{2017}]{2017arXiv170607478O}
{Oman} K.~A.,  {Marasco} A.,  {Navarro} J.~F.,  {Frenk} C.~S.,  {Schaye} J.,
  {Ben{\'{\i}}tez-Llambay} A.,  2017, preprint, \href
  {http://adsabs.harvard.edu/abs/2017arXiv170607478O} {} (\mn@eprint {arXiv}
  {1706.07478})

\bibitem[\protect\citeauthoryear{{Pace}}{{Pace}}{2016}]{2016PhDT.......110P}
{Pace} A.~B.,  2016, PhD thesis, University of California, Irvine

\bibitem[\protect\citeauthoryear{{Pe{\~n}arrubia}, {Navarro}  \&
  {McConnachie}}{{Pe{\~n}arrubia} et~al.}{2008}]{2008ApJ...673..226P}
{Pe{\~n}arrubia} J.,  {Navarro} J.~F.,   {McConnachie} A.~W.,  2008, \mn@doi
  [\apj] {10.1086/523686}, \href
  {http://adsabs.harvard.edu/abs/2008ApJ...673..226P} {673, 226}

\bibitem[\protect\citeauthoryear{{Peng} et~al.,}{{Peng}
  et~al.}{2008}]{2008ApJ...681..197P}
{Peng} E.~W.,  et~al., 2008, \mn@doi [\apj] {10.1086/587951}, \href
  {http://adsabs.harvard.edu/abs/2008ApJ...681..197P} {681, 197}

\bibitem[\protect\citeauthoryear{{Read} \& {Erkal}}{{Read} \&
  {Erkal}}{2018}]{2018arXiv180707093R}
{Read} J.~I.,  {Erkal} D.,  2018, preprint, \href
  {http://adsabs.harvard.edu/abs/2018arXiv180707093R} {} (\mn@eprint {arXiv}
  {1807.07093})

\bibitem[\protect\citeauthoryear{{Read} \& {Gilmore}}{{Read} \&
  {Gilmore}}{2005}]{2005MNRAS.356..107R}
{Read} J.~I.,  {Gilmore} G.,  2005, \mnras, \href
  {http://adsabs.harvard.edu/cgi-bin/nph-bib_query?bibcode=2005MNRAS.356..107R&db_key=AST}
  {356, 107}

\bibitem[\protect\citeauthoryear{{Read} \& {Trentham}}{{Read} \&
  {Trentham}}{2005}]{2005RSPTA.363.2693R}
{Read} J.~I.,  {Trentham} N.,  2005, \mn@doi [Philosophical Transactions of the
  Royal Society of London Series A] {10.1098/rsta.2005.1648}, \href
  {http://adsabs.harvard.edu/abs/2005RSPTA.363.2693R} {363}

\bibitem[\protect\citeauthoryear{{Read}, {Wilkinson}, {Evans}, {Gilmore}  \&
  {Kleyna}}{{Read} et~al.}{2006a}]{2006MNRAS.366..429R}
{Read} J.~I.,  {Wilkinson} M.~I.,  {Evans} N.~W.,  {Gilmore} G.,   {Kleyna}
  J.~T.,  2006a, \mn@doi [\mnras] {10.1111/j.1365-2966.2005.09861.x}, \href
  {http://adsabs.harvard.edu/abs/2006MNRAS.366..429R} {366, 429}

\bibitem[\protect\citeauthoryear{{Read}, {Wilkinson}, {Evans}, {Gilmore}  \&
  {Kleyna}}{{Read} et~al.}{2006b}]{2006MNRAS.367..387R}
{Read} J.~I.,  {Wilkinson} M.~I.,  {Evans} N.~W.,  {Gilmore} G.,   {Kleyna}
  J.~T.,  2006b, \mn@doi [\mnras] {10.1111/j.1365-2966.2005.09959.x}, \href
  {http://adsabs.harvard.edu/abs/2006MNRAS.367..387R} {367, 387}

\bibitem[\protect\citeauthoryear{{Read}, {Pontzen}  \& {Viel}}{{Read}
  et~al.}{2006c}]{2006MNRAS.371..885R}
{Read} J.~I.,  {Pontzen} A.~P.,   {Viel} M.,  2006c, \mn@doi [\mnras]
  {10.1111/j.1365-2966.2006.10720.x}, \href
  {http://adsabs.harvard.edu/abs/2006MNRAS.371..885R} {371, 885}

\bibitem[\protect\citeauthoryear{{Read}, {Agertz}  \& {Collins}}{{Read}
  et~al.}{2016a}]{Read+16}
{Read} J.~I.,  {Agertz} O.,   {Collins} M.~L.~M.,  2016a, \mn@doi [\mnras]
  {10.1093/mnras/stw713}, \href
  {http://adsabs.harvard.edu/abs/2016MNRAS.459.2573R} {459, 2573}

\bibitem[\protect\citeauthoryear{{Read}, {Iorio}, {Agertz}  \&
  {Fraternali}}{{Read} et~al.}{2016b}]{2016MNRAS.462.3628R}
{Read} J.~I.,  {Iorio} G.,  {Agertz} O.,   {Fraternali} F.,  2016b, \mn@doi
  [\mnras] {10.1093/mnras/stw1876}, \href
  {http://adsabs.harvard.edu/abs/2016MNRAS.462.3628R} {462, 3628}

\bibitem[\protect\citeauthoryear{{Read}, {Iorio}, {Agertz}  \&
  {Fraternali}}{{Read} et~al.}{2017}]{2017MNRAS.467.2019R}
{Read} J.~I.,  {Iorio} G.,  {Agertz} O.,   {Fraternali} F.,  2017, \mn@doi
  [\mnras] {10.1093/mnras/stx147}, \href
  {http://esoads.eso.org/abs/2017MNRAS.467.2019R} {467, 2019}

\bibitem[\protect\citeauthoryear{{Rodr{\'{\i}}guez-Puebla}, {Primack},
  {Avila-Reese}  \& {Faber}}{{Rodr{\'{\i}}guez-Puebla}
  et~al.}{2017}]{2017MNRAS.470..651R}
{Rodr{\'{\i}}guez-Puebla} A.,  {Primack} J.~R.,  {Avila-Reese} V.,   {Faber}
  S.~M.,  2017, \mn@doi [\mnras] {10.1093/mnras/stx1172}, \href
  {http://adsabs.harvard.edu/abs/2017MNRAS.470..651R} {470, 651}

\bibitem[\protect\citeauthoryear{{Ryan-Weber}, {Begum}, {Oosterloo}, {Pal},
  {Irwin}, {Belokurov}, {Evans}  \& {Zucker}}{{Ryan-Weber}
  et~al.}{2008}]{2008MNRAS.384..535R}
{Ryan-Weber} E.~V.,  {Begum} A.,  {Oosterloo} T.,  {Pal} S.,  {Irwin} M.~J.,
  {Belokurov} V.,  {Evans} N.~W.,   {Zucker} D.~B.,  2008, \mn@doi [\mnras]
  {10.1111/j.1365-2966.2007.12734.x}, \href
  {http://adsabs.harvard.edu/abs/2008MNRAS.384..535R} {384, 535}

\bibitem[\protect\citeauthoryear{{Sanders}, {Evans}  \& {Dehnen}}{{Sanders}
  et~al.}{2018}]{2018MNRAS.478.3879S}
{Sanders} J.~L.,  {Evans} N.~W.,   {Dehnen} W.,  2018, \mn@doi [\mnras]
  {10.1093/mnras/sty1278}, \href
  {http://adsabs.harvard.edu/abs/2018MNRAS.478.3879S} {478, 3879}

\bibitem[\protect\citeauthoryear{{Sawala} et~al.,}{{Sawala}
  et~al.}{2015}]{2015MNRAS.448.2941S}
{Sawala} T.,  et~al., 2015, \mn@doi [\mnras] {10.1093/mnras/stu2753}, \href
  {http://adsabs.harvard.edu/abs/2015MNRAS.448.2941S} {448, 2941}

\bibitem[\protect\citeauthoryear{{Schneider}, {Trujillo-Gomez}, {Papastergis},
  {Reed}  \& {Lake}}{{Schneider} et~al.}{2017}]{2017MNRAS.470.1542S}
{Schneider} A.,  {Trujillo-Gomez} S.,  {Papastergis} E.,  {Reed} D.~S.,
  {Lake} G.,  2017, \mn@doi [\mnras] {10.1093/mnras/stx1294}, \href
  {http://adsabs.harvard.edu/abs/2017MNRAS.470.1542S} {470, 1542}

\bibitem[\protect\citeauthoryear{{Shen}, {Madau}, {Conroy}, {Governato}  \&
  {Mayer}}{{Shen} et~al.}{2014}]{2014ApJ...792...99S}
{Shen} S.,  {Madau} P.,  {Conroy} C.,  {Governato} F.,   {Mayer} L.,  2014,
  \mn@doi [\apj] {10.1088/0004-637X/792/2/99}, \href
  {http://adsabs.harvard.edu/abs/2014ApJ...792...99S} {792, 99}

\bibitem[\protect\citeauthoryear{{Sigad}, {Kolatt}, {Bullock}, {Kravtsov},
  {Klypin}, {Primack}  \& {Dekel}}{{Sigad} et~al.}{2000}]{2000astro.ph..5323S}
{Sigad} Y.,  {Kolatt} T.~S.,  {Bullock} J.~S.,  {Kravtsov} A.~V.,  {Klypin}
  A.~A.,  {Primack} J.~R.,   {Dekel} A.,  2000, ArXiv Astrophysics e-prints,
  \href {http://adsabs.harvard.edu/abs/2000astro.ph..5323S} {}

\bibitem[\protect\citeauthoryear{{Sohn}, {Watkins}, {Fardal}, {van der Marel},
  {Deason}, {Besla}  \& {Bellini}}{{Sohn} et~al.}{2018}]{2018arXiv180401994S}
{Sohn} S.~T.,  {Watkins} L.~L.,  {Fardal} M.~A.,  {van der Marel} R.~P.,
  {Deason} A.~J.,  {Besla} G.,   {Bellini} A.,  2018, preprint, \href
  {http://adsabs.harvard.edu/abs/2018arXiv180401994S} {} (\mn@eprint {arXiv}
  {1804.01994})

\bibitem[\protect\citeauthoryear{{Spergel} \& {Steinhardt}}{{Spergel} \&
  {Steinhardt}}{2000}]{2000PhRvL..84.3760S}
{Spergel} D.~N.,  {Steinhardt} P.~J.,  2000, \mn@doi [Physical Review Letters]
  {10.1103/PhysRevLett.84.3760}, \href
  {http://adsabs.harvard.edu/abs/2000PhRvL..84.3760S} {84, 3760}

\bibitem[\protect\citeauthoryear{{Spitler} \& {Forbes}}{{Spitler} \&
  {Forbes}}{2009}]{2009MNRAS.392L...1S}
{Spitler} L.~R.,  {Forbes} D.~A.,  2009, \mn@doi [\mnras]
  {10.1111/j.1745-3933.2008.00567.x}, \href
  {http://adsabs.harvard.edu/abs/2009MNRAS.392L...1S} {392, L1}

\bibitem[\protect\citeauthoryear{{Stanimirovi{\'c}}, {Staveley-Smith}  \&
  {Jones}}{{Stanimirovi{\'c}} et~al.}{2004}]{2004ApJ...604..176S}
{Stanimirovi{\'c}} S.,  {Staveley-Smith} L.,   {Jones} P.~A.,  2004, \mn@doi
  [\apj] {10.1086/381869}, \href
  {http://adsabs.harvard.edu/abs/2004ApJ...604..176S} {604, 176}

\bibitem[\protect\citeauthoryear{{Stephens}, {Catelan}  \&
  {Contreras}}{{Stephens} et~al.}{2006}]{2006AJ....131.1426S}
{Stephens} A.~W.,  {Catelan} M.,   {Contreras} R.~P.,  2006, \mn@doi [\aj]
  {10.1086/500300}, \href {http://adsabs.harvard.edu/abs/2006AJ....131.1426S}
  {131, 1426}

\bibitem[\protect\citeauthoryear{{Strader}, {Smith}, {Larsen}, {Brodie}  \&
  {Huchra}}{{Strader} et~al.}{2009}]{2009AJ....138..547S}
{Strader} J.,  {Smith} G.~H.,  {Larsen} S.,  {Brodie} J.~P.,   {Huchra} J.~P.,
  2009, \mn@doi [\aj] {10.1088/0004-6256/138/2/547}, \href
  {http://adsabs.harvard.edu/abs/2009AJ....138..547S} {138, 547}

\bibitem[\protect\citeauthoryear{{Torrealba}, {Koposov}, {Belokurov}  \&
  {Irwin}}{{Torrealba} et~al.}{2016}]{2016MNRAS.459.2370T}
{Torrealba} G.,  {Koposov} S.~E.,  {Belokurov} V.,   {Irwin} M.,  2016, \mn@doi
  [\mnras] {10.1093/mnras/stw733}, \href
  {http://adsabs.harvard.edu/abs/2016MNRAS.459.2370T} {459, 2370}

\bibitem[\protect\citeauthoryear{{Trujillo} et~al.,}{{Trujillo}
  et~al.}{2018}]{2018arXiv180610141T}
{Trujillo} I.,  et~al., 2018, preprint, \href
  {http://adsabs.harvard.edu/abs/2018arXiv180610141T} {} (\mn@eprint {arXiv}
  {1806.10141})

\bibitem[\protect\citeauthoryear{{Ural}, {Wilkinson}, {Read}  \&
  {Walker}}{{Ural} et~al.}{2015}]{2015NatCo...6E7599U}
{Ural} U.,  {Wilkinson} M.~I.,  {Read} J.~I.,   {Walker} M.~G.,  2015, \mn@doi
  [Nature Communications] {10.1038/ncomms8599}, \href
  {http://adsabs.harvard.edu/abs/2015NatCo...6E7599U} {6, 7599}

\bibitem[\protect\citeauthoryear{{Veljanoski} et~al.,}{{Veljanoski}
  et~al.}{2013}]{2013MNRAS.435.3654V}
{Veljanoski} J.,  et~al., 2013, \mn@doi [\mnras] {10.1093/mnras/stt1557}, \href
  {http://adsabs.harvard.edu/abs/2013MNRAS.435.3654V} {435, 3654}

\bibitem[\protect\citeauthoryear{{Veljanoski} et~al.,}{{Veljanoski}
  et~al.}{2014}]{2014MNRAS.442.2929V}
{Veljanoski} J.,  et~al., 2014, \mn@doi [\mnras] {10.1093/mnras/stu1055}, \href
  {http://adsabs.harvard.edu/abs/2014MNRAS.442.2929V} {442, 2929}

\bibitem[\protect\citeauthoryear{{Veljanoski} et~al.,}{{Veljanoski}
  et~al.}{2015}]{2015MNRAS.452..320V}
{Veljanoski} J.,  et~al., 2015, \mn@doi [\mnras] {10.1093/mnras/stv1259}, \href
  {http://adsabs.harvard.edu/abs/2015MNRAS.452..320V} {452, 320}

\bibitem[\protect\citeauthoryear{{Walker}, {Mateo}  \& {Olszewski}}{{Walker}
  et~al.}{2009a}]{2009AJ....137.3100W}
{Walker} M.~G.,  {Mateo} M.,   {Olszewski} E.~W.,  2009a, \mn@doi [\aj]
  {10.1088/0004-6256/137/2/3100}, \href
  {http://adsabs.harvard.edu/abs/2009AJ....137.3100W} {137, 3100}

\bibitem[\protect\citeauthoryear{{Walker}, {Mateo}, {Olszewski},
  {Pe{\~n}arrubia}, {Wyn Evans}  \& {Gilmore}}{{Walker}
  et~al.}{2009b}]{2009ApJ...704.1274W}
{Walker} M.~G.,  {Mateo} M.,  {Olszewski} E.~W.,  {Pe{\~n}arrubia} J.,  {Wyn
  Evans} N.,   {Gilmore} G.,  2009b, \mn@doi [\apj]
  {10.1088/0004-637X/704/2/1274}, \href
  {http://adsabs.harvard.edu/abs/2009ApJ...704.1274W} {704, 1274}

\bibitem[\protect\citeauthoryear{{Weisz}, {Dolphin}, {Skillman}, {Holtzman},
  {Gilbert}, {Dalcanton}  \& {Williams}}{{Weisz}
  et~al.}{2014}]{2014ApJ...789..147W}
{Weisz} D.~R.,  {Dolphin} A.~E.,  {Skillman} E.~D.,  {Holtzman} J.,  {Gilbert}
  K.~M.,  {Dalcanton} J.~J.,   {Williams} B.~F.,  2014, \mn@doi [\apj]
  {10.1088/0004-637X/789/2/147}, \href
  {http://adsabs.harvard.edu/abs/2014ApJ...789..147W} {789, 147}

\bibitem[\protect\citeauthoryear{{Weisz} et~al.,}{{Weisz}
  et~al.}{2016}]{2016ApJ...822...32W}
{Weisz} D.~R.,  et~al., 2016, \mn@doi [\apj] {10.3847/0004-637X/822/1/32},
  \href {http://adsabs.harvard.edu/abs/2016ApJ...822...32W} {822, 32}

\bibitem[\protect\citeauthoryear{{Wheeler}, {O{\~n}orbe}, {Bullock},
  {Boylan-Kolchin}, {Elbert}, {Garrison-Kimmel}, {Hopkins}  \& {Kere{\v
  s}}}{{Wheeler} et~al.}{2015}]{2015MNRAS.453.1305W}
{Wheeler} C.,  {O{\~n}orbe} J.,  {Bullock} J.~S.,  {Boylan-Kolchin} M.,
  {Elbert} O.~D.,  {Garrison-Kimmel} S.,  {Hopkins} P.~F.,   {Kere{\v s}} D.,
  2015, \mn@doi [\mnras] {10.1093/mnras/stv1691}, \href
  {http://adsabs.harvard.edu/abs/2015MNRAS.453.1305W} {453, 1305}

\bibitem[\protect\citeauthoryear{{Zaritsky}, {Crnojevi{\'c}}  \&
  {Sand}}{{Zaritsky} et~al.}{2016}]{2016ApJ...826L...9Z}
{Zaritsky} D.,  {Crnojevi{\'c}} D.,   {Sand} D.~J.,  2016, \mn@doi [\apjl]
  {10.3847/2041-8205/826/1/L9}, \href
  {http://adsabs.harvard.edu/abs/2016ApJ...826L...9Z} {826, L9}

\bibitem[\protect\citeauthoryear{{van Dokkum} et~al.,}{{van Dokkum}
  et~al.}{2016}]{2016ApJ...828L...6V}
{van Dokkum} P.,  et~al., 2016, \mn@doi [\apjl] {10.3847/2041-8205/828/1/L6},
  \href {http://adsabs.harvard.edu/abs/2016ApJ...828L...6V} {828, L6}

\bibitem[\protect\citeauthoryear{{van Dokkum} et~al.,}{{van Dokkum}
  et~al.}{2017}]{2017ApJ...844L..11V}
{van Dokkum} P.,  et~al., 2017, \mn@doi [\apjl] {10.3847/2041-8213/aa7ca2},
  \href {http://adsabs.harvard.edu/abs/2017ApJ...844L..11V} {844, L11}

\bibitem[\protect\citeauthoryear{{van Dokkum}, {Danieli}, {Cohen}  \&
  {Conroy}}{{van Dokkum} et~al.}{2018a}]{2018arXiv180706025V}
{van Dokkum} P.,  {Danieli} S.,  {Cohen} Y.,   {Conroy} C.,  2018a, preprint,
  \href {http://adsabs.harvard.edu/abs/2018arXiv180706025V} {} (\mn@eprint
  {arXiv} {1807.06025})

\bibitem[\protect\citeauthoryear{{van Dokkum} et~al.,}{{van Dokkum}
  et~al.}{2018b}]{2018Natur.555..629V}
{van Dokkum} P.,  et~al., 2018b, \mn@doi [\nat] {10.1038/nature25767}, \href
  {http://adsabs.harvard.edu/abs/2018Natur.555..629V} {555, 629}

\bibitem[\protect\citeauthoryear{{van der Marel} \& {Sahlmann}}{{van der Marel}
  \& {Sahlmann}}{2016}]{2016ApJ...832L..23V}
{van der Marel} R.~P.,  {Sahlmann} J.,  2016, \mn@doi [\apjl]
  {10.3847/2041-8205/832/2/L23}, \href
  {http://adsabs.harvard.edu/abs/2016ApJ...832L..23V} {832, L23}

\makeatother
\end{thebibliography}



\appendix


\section{Galaxies with globular cluster systems but lacking kinematic information}

Below we list a handful of  interesting nearby dwarf galaxies that possess at least one GC but lack a stellar velocity dispersion or an HI rotation measure in order to estimate their total halo mass. The basic properties of the host galaxy, along with the total mass in the GC system are listed in Table 4. 

\subsection{Phoenix}
\citet{1999AJ....118..862M}
identified 4 possible low luminosity GCs in the Local Group Phoenix galaxy. They have luminosities of 
$M_V$ = --4.39, --4.14, --4.02, --4.24, and a total GC system mass of $3.1\times10^4$ $\Msun$. 
We note that the HI listed by \citet{2012AJ....144....4M} for Phoenix is in the form of spatially offset HI clouds and so their association with the galaxy is uncertain. 


\subsection{KKs3 (PGC 009140)}
This isolated dwarf has 1 GC ($M_V = -8.33$, $M_{\ast} = 3.45 \times 10^5\, \Msun$) discovered in HST/ACS imaging by \citet{2015AN....336..707K}.
The galaxy is located 2 Mpc just outside of the Local Group, with $M_B$ = --10.8 ($M_{\ast}$ $\sim$ $4.5\times10^6\,\Msun$). Its sole GC represents almost 8\% of the galaxy stellar mass. It has some HI gas but no HI rotation curve reported. 



\subsection{IKN}
The IKN dSph is similar to the Fornax dwarf with 5 GCs (their total mass  is  $8.0\times10^5$ $\Msun$). At a distance of 3.8 Mpc, the galaxy has a luminosity of $M_V = -11.5$, giving it a higher GC specific frequency than Fornax or equivalently the GCs represent about 10\% of the galaxy stellar mass. \citet{2014A&A...565A..98L}
presents spectra of the GCs. 
The galaxy does not appear to possess any HI gas 
\citep{2010MNRAS.406.1967G}. There also appears to be no central velocity dispersion or half-light radius in the literature. This is probably due to the presence of a nearby, very bright star.

\section{Galaxies lacking a globular cluster system}

Below we list some interesting nearby dwarf galaxies that lack a reported GC system. Some of the galaxies listed below either have a significant halo mass that suggests some GCs could have been formed. Others contain HI gas -- the presence of which suggests that the host galaxy has not undergone ram pressure stripping or a tidal interaction. The lack of tidal interaction suggests any GCs formed have not been subsequently removed. Placing good limits on the absence of a GC system in a galaxy with a substantial halo mass and the presence of HI gas would provide key information on the stochastic nature of GC formation in low mass galaxies. 



\subsection{Carina}
The Carina dSph has no known GCs. It lies 107 kpc from the Milky Way, with $M_V = -9.2$ ($4\times10^5\,\Msun$). No HI has been detected according to \citet{2012AJ....144....4M}. The halo mass of Carina has been modelled by \cite{2015NatCo...6E7599U} 
to be log $M_{200}$ = 8.56 and, based on its observed HI rotation curve, to be 8.55 by \citet{2017MNRAS.467.2019R}. 
Based on this halo mass we expect $\le$1 GC.

\subsection{LeoT}
The LeoT galaxy lies just beyond the virial radius of the Milky Way at 410 kpc. It has a luminosity of 
$M_V$ = --8 and HI gas mass of $2.8\times10^5$ $\Msun$ \citet{2008MNRAS.384..535R} 
It has $\Reff$ = 178 pc and $\sigma$ = 7.5 km/s and a halo mass quoted by \citet{2017MNRAS.467.2019R} 
of log $M_{200}$ = 8.54--8.88. For this halo mass we expect $\le$1 GC, thus the apparent lack of a GC is not surprising. 



\subsection{IC 10 and IC 1613}
IC 10 ($M_V = -15.0$) and IC 1613 ($M_V = -15.2$) appear to be the highest luminosity Local Group galaxies (excluding special case of M32) without a single GC. 
We note that a number (66) of star clusters have been catalogued in IC 10 by \citet{2015ApJ...804..123L} 
but it is not clear if they are old GC-like. 
Both galaxies reveal the presence of HI gas \citep{2012AJ....144....4M} and have halo masses estimated from HI rotation curves of log  $M_{200}$ = 10.2 for IC 10 and 8.8 for IC 1613  \citep{2015AJ....149..180O}. 
\citet{2017MNRAS.467.2019R} 
find a wide range of halo mass, i.e. log $M_{200}$ = 8.85--10.07. Based on these halo mass estimates we would expect IC 10 and IC 1613 to host several GCs.

\subsection{Sextens A and B} 
Sextens A has a number of star clusters in its main body but it seems that none of them are GCs \citep{2014A&A...566A..44B}. 
However, one star cluster outside the main body appears to be a bona-fide GC (Beasley 2018, priv. comm.). 
\citet{2014AstBu..69....1M} report the discovery of a young ($\sim$2 Gyr) compact star cluster in Sextens B using WFPC2 imaging. The star cluster is quite luminous with $M_V = -7.77$. Given its young age, we do not classify it as a GC. Only Sextens B has a listed velocity dispersion, i.e. $\sigma$ = 8.0 km/s and half-light radius of 440 pc. Both galaxies reveal HI gas \citet{2012AJ....144....4M} so further study of the existing star clusters are warranted. Both galaxies lie just beyond the `zero velocity surface' of the Local Group. 

\subsection{LGS3, Leo~A, SAG dIrr, UGC 4879 and Antlia} 
These Local Group dwarfs are located at large distances from the Milky Way and M31 and are listed by \citet{2012AJ....144....4M} as containing HI gas. This suggests that they have not undergone a strong tidal interaction with either of the giant galaxies and so any GCs that they formed should remain. We note that LGS3 and SAG dIrr are part of the Little Things survey but were not included in the high resolution HI observations of \citet{2015AJ....149..180O}.

\subsection{Crater~2}
In 2014, the discovery of the distant (145 kpc) GC Crater was announced by \citet{2014MNRAS.441.2124B} and \cite{2014ApJ...786L...3L}. 
Follow-up HST/ACS imaging by \citet{2016ApJ...822...32W}
derived an age of 7.5 Gyr, [M/H] = --1.65 and $M_V$ = -5.3 (and hence a stellar mass of $2.1 \times 10^4 \Msun$, assuming $M/L_V$ = 1.88). Two years after the discovery of the Crater GC, it was suggested that it is associated with the newly discovered Crater 2 galaxy by \citet{2016MNRAS.459.2370T}. 
They argued that the galaxy Crater 2, and the GC Crater, share the same great circle (and hence a physical association) that includes the Leo dwarf galaxies. The Crater 2 galaxy has a luminosity of $M_V = -8.2 \pm 0.1$ (or stellar mass of $\sim1.6\times10^5$\,M$_{\odot}$, assuming $M/L_V$ = 1).  It has a half light radius of 1066 pc and $\sigma$ = 2.7 km/s. 
Currently, the association of the Crater GC with the Crater 2 galaxy is highly uncertain and the galaxy appears to have undergone severe ($\sim$90) per cent mass loss due to tidal disruption \citep{2018MNRAS.478.3879S}. 
We have decided not to 
include Crater 2 in the main part of this paper. 

\subsection{And XVII}
\citet{2008ApJ...676L..17I}
suggested that And XVII may have associated GCs. Based on their velocities, \citet{2014MNRAS.442.2929V} argued against this and associated the GCs with M31. We do not include And XVII in Table 1. 

\section{GC System Number--Halo Mass Relation}

\begin{figure}
	\includegraphics[width=7.2cm, angle=-90]{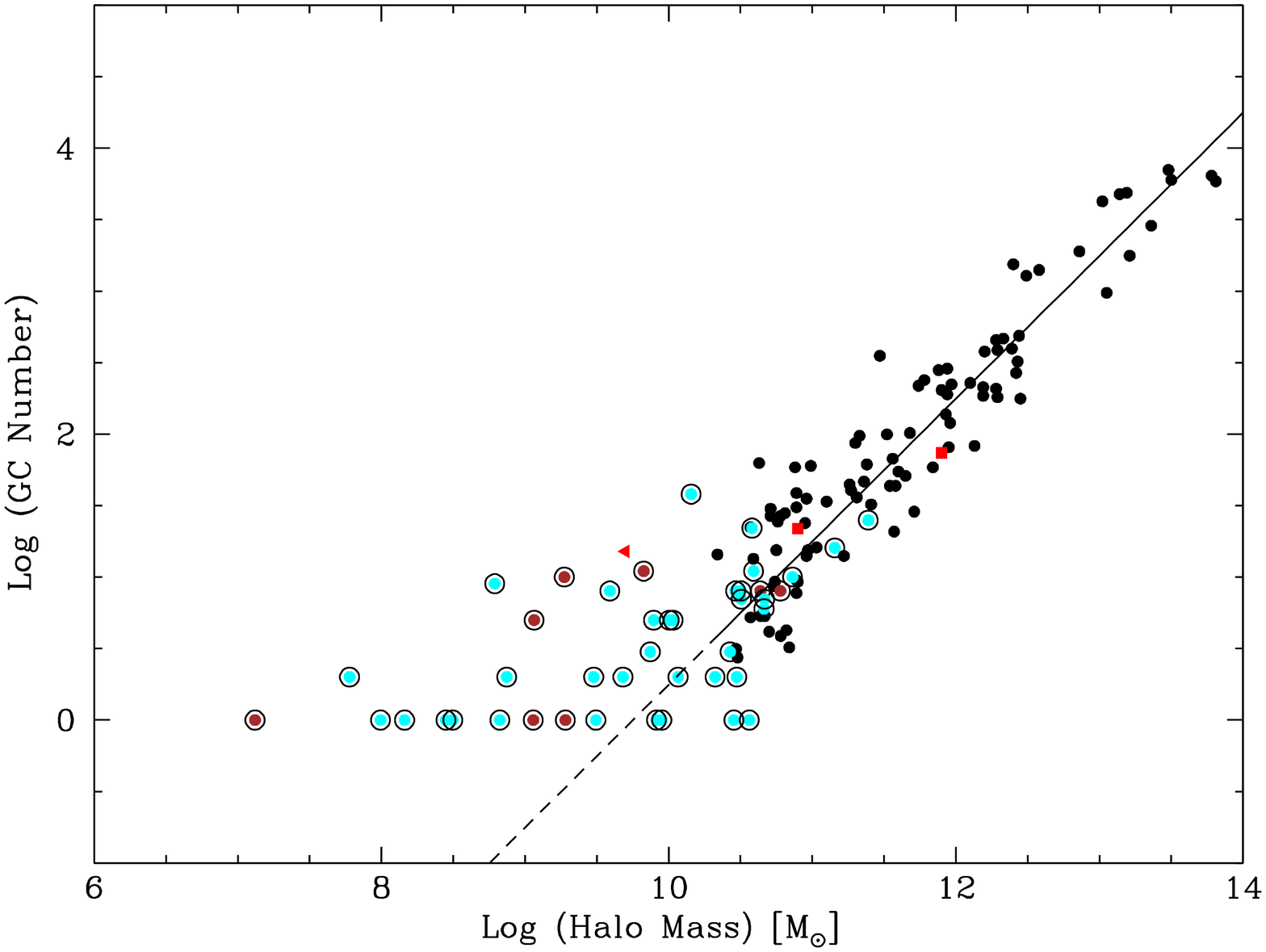}
    \caption{Globular cluster system number-halo mass relation. Same symbols used as in Fig.~\ref{fig:halo}. Although many low mass galaxies follow the relation from higher masses, many others have GCs of lower than average mean mass and hence deviate from the extrapolated relation. 
    }
    \label{fig:num}
\end{figure}

In Fig.~\ref{fig:num} we show the number of GCs in a galaxy vs the galaxy halo mass. For the high mass galaxies from Spitler \& Forbes (2009) we simply divide the reported GC system mass by the mean mass of a GC assumed in that work, i.e. 4 $\times$ 10$^5$ $M_{\odot}$. For the low mass galaxies we use the number of GCs quoted in Tables 1 and 2.

The plot shows a tight relation when the GC system exceeds 10 GCs, however below that number the relation shows a strong deviation from the linear relation. This deviation is, at least in part, due to the breakdown of the assumption of a common mean GC mass. Low mass galaxies have, on average, lower mass GCs. Hence the importance in this work to coadd  individual GC masses to calculate a total GC system mass and thus test whether that quantity continues to have a linear relationship with halo mass.

Fig~\ref{fig:num} also shows that some galaxies with very small GC systems, even some with a single GC, still follow the GC number vs halo mass relation. This illustrates that even a single GC in a low mass galaxy can, in some cases, have a 'typical' GC mass. Whether this is a survival or a formation feature is unknown. The UDGs DF44 and VCC1287 lie within the scatter of the main relation.

\bsp	
\label{lastpage}
\end{document}